\documentclass[aps,prb,floats,twocolumn]{revtex4-1}
\usepackage{graphicx}
\usepackage{amssymb}
\usepackage{amsmath}
\usepackage{color}
\usepackage[dvipsnames]{xcolor}
\usepackage{float}
\graphicspath{{./FIG_FINALI/}}
\usepackage[caption=false]{subfig}
\usepackage{array}
\usepackage{braket}
\usepackage{scrextend}

\begin{document}
%\title{Spin-orbit Hamiltonians For Organic Crystals In Terms of Wannier Functions from First Principles Calculation}
\title{Spin-orbit Hamiltonian for organic crystals from first principles electronic structure and Wannier functions}
\author{Subhayan Roychoudhury}
\affiliation{School of Physics, AMBER and CRANN Institute, Trinity College Dublin, Dublin 2, Ireland}
\author{Stefano Sanvito}
\affiliation{School of Physics, AMBER and CRANN Institute, Trinity College Dublin, Dublin 2, Ireland}

\begin{abstract}
Spin-orbit coupling in organic crystals is responsible for many spin-relaxation phenomena, going from
spin diffusion to intersystem crossing. With the goal of constructing effective spin-orbit Hamiltonians
to be used in multiscale approaches to the thermodynamical properties of organic crystals, we present 
a method that combines density functional theory with the construction of Wannier functions. In particular
we show that the spin-orbit Hamiltonian constructed over maximally localised Wannier functions can be 
computed by direct evaluation of the spin-orbit matrix elements over the Wannier functions constructed 
in absence of spin-orbit interaction. This eliminates the problem of computing the Wannier functions for 
almost degenerate bands, a problem always present with the spin-orbit-split bands of organic crystals.
Examples of the method are presented for isolated organic molecules, for mono-dimensional chains of 
Pb and C atoms and for triarylamine-based one-dimansional single crystals.
\end{abstract}

%\date{}
\maketitle
%\section{}
%\subsection}{

\section{Introduction}

Spintronics devices operate by detecting the spin of a carrier in the same way as a regular electronic device 
measures its electrical charge~\cite{Wolf1488}. These devices are already the state of the art in the design of
magnetic sensors such as the magnetic read-head of hard-disk drives~\cite{Ornes05032013}, but also have 
excellent prospect as logic gate elements~\cite{DattaSLogic,1995PhT,Awschalom,Igor}. Logic circuits using the 
spin degree of freedom may offer low energy consumption and high speed owing to the fact that the dynamics of 
spins takes place at a much smaller energy scale than that of the charge~\cite{DattaSLogic,Wolf1488}. 

Recent years have also witnessed a marked increase in interest into investigations of organic molecules and 
molecular crystals as materials platform, initially for electronics~\cite{HoroGill,JOUR1} and lately also for 
spintronics~\cite{Dediu2002181,JOUR,C1CS15047B}. The main reason behind such interest is that organic 
crystals, coming in a wide chemical variety, are typically much more flexible than their inorganic counterparts 
and they can exhibit an ample range of electronic properties, which are highly tuneable in practice. For example, 
it is possible to change the conductivity of organic polymers over fifteen orders of magnitude~\cite{PhysRevLett.39.1098}. 
In addition to such extreme spectrum of physical/chemical properties organic materials are usually processed at 
low temperature. This is an advantage over inorganic compounds, which translates into a drastic reduction of the 
typical manufacturing and infrastructure costs~\cite{Forrest}. Finally, specific to spintronics is the fact that 
both the spin-orbit (SO) and hyperfine interaction are very weak~\cite{Sanvito} in organic compounds, resulting 
in a weak spin scattering during the electron transport~\cite{Pramanik, Tsukagoshi,SCS2009}. 

Regardless of the type of media used, either organic or inorganic, spintronics always concerns phenomena related to 
the injection, manipulation and detection of spins into a solid state environment~\cite{C1CS15047B}. In the prototypical 
spintronic device, the spin-valve~\cite{SpinValve}, a non-magnetic spacer is sandwiched between 
two ferromagnents. Spins, which are initially aligned along the magnetization vector of the first ferromagnet, travel to 
the other ferromagnent through the spacer, and the resistance of the entire device depends on the relative orientation 
of the magnetization vectors of the two magnets. However, if the spin direction is lost across the spacer, the resistance 
will become independent of the magnetic configuration of the device. As such, in order to measure any spin-dependent 
effect one has to ensure that the charge carriers maintain their spin direction through the spacer. Notably, this requirement 
is not only demanded by spin-valves, but also by any devices based on spins. There are several mechanism for spin-relaxation 
in the solid state~\cite{RevModPhys.76.323}.

In an organic semiconductor (OSC) the unwanted spin-relaxation can be caused by the presence of paramagnetic 
impurities, by SO coupling and by hyperfine interaction. In general paramagnetic impurities can be controlled to a very 
high degree of precision and they can be almost completely eliminated from an OSC during the chemical 
synthesis~\cite{Impurity}.
The hyperfine interaction instead can be usually considered small. This is because there are only a few elements 
typically present in organic molecules with abundant isotopes baring nuclear spins. The most obvious exception is 
hydrogen. However, most of the OSC crystals are $\pi$-conjugated and the $\pi$-states, responsible for the extremal 
energy levels, and hence for the electron transport, are usually delocalized. This means that the overlap of the wave 
function over the H nuclei has to be considered small. Finally, also the SO coupling is weak owing to the fact that most 
of the atoms composing organic compounds are light. 

As such, since all the non-spin-conserving interactions are weak
in OSCs, it is not surprising that there is contradictory evidence concerning the interaction mostly responsible for spin-diffusion
in organic crystals. Conflicting experimental evidence exists supporting either the SO coupling~\cite{PhysRevB.81.153202,Drew} 
or the hyperfine interaction~\cite{PhysRevB.75.245324,PhysRevB.78.115203}, indicating that the dominant mechanism 
may depend on the specific material under investigation. For this reason it is important to develop methods for determining 
the strength of both the SO and the hyperfine coupling in real materials. These can eventually be the basis for constructing 
effective Hamiltonians to be used for the evaluation of the relevant thermodynamics quantities (e.g. the spin diffusion length). 
Here we present one of such methods for the case of the SO interaction.

The SO interaction is a relativistic effect arising from the electron motion in the nuclear potential. In the electron reference 
frame the nucleus moves and creates a magnetic field, which in turn interacts with the electron spin. This is the spin-orbit 
coupling~\cite{cohen1977quantum}. Since the SO interaction allows the spin of an electron to change direction during the 
electron motion, it is an interaction responsible for spin relaxation. In fact, there exist several SO-based microscopic theories 
of spin relaxation in solid state systems~\cite{RevModPhys.76.323}. In the case of inorganic semiconductors these usually 
require knowledge of the band-structure of the material, some information about its mobility and an estimate of the spin-orbit
strength. In the case of OSCs the situation, however, is more complex, mostly because the transport mechanism is more
difficult to describe. Firstly, the band picture holds true only for a few cases, while for many others one has to consider the 
material as an ensemble of weakly coupled molecules with a broad distribution of hopping integrals~\cite{Troisi}. 
Secondly, the typical phonon energies are of the same order of magnitude of the electronic band width, indicating that 
electron-phonon scattering cannot be treated as a perturbation of the band structure. For all these reasons the description 
of the thermodynamical properties of OSCs requires the construction of a multi-scale theory, where the elementary electronic 
structure is mapped onto an effective Hamiltonian retaining only a handful of the original degrees of freedom~\cite{doi:10.1021/ct500390a}. 
A rigorous and now standard method for constructing such effective Hamiltonian consists in calculating the band structure 
over a set of Wannier functions~\cite{PhysRev.52.191,RevModPhys.34.645}. These can be constructed in a very general 
way as the Fourier transform of a linear combination of Bloch states, where the linear combination is taken so to minimize 
the spatial extension of the Wannier functions. These are the so-called maximally localized Wannier fuctions 
(MLWFs)~\cite{PhysRevB.56.12847,RevModPhys.84.1419}. 

The MLWF method performs best for well-isolated bands. This is indeed the case of OSCs, where often the valence and 
conduction bands originate respectively from the highest occupied molecular orbital (HOMO) and the lowest unoccupied
molecular orbital (LUMO) of the gas-phase molecule. In fact, when the MLWF procedure is applied to such band structure 
one obtains Wannier orbitals almost identical to the molecule HOMO and LUMO~\cite{doi:10.1021/ct500390a}. Spin-orbit interaction, 
however, splits such well-defined bands, and in OSCs the split is typically a few tenths of $\mu$eV. Thus, in this case, one 
has to apply the MLWF procedure to bands, which are indistinguishable at an energy scale larger then a few $\mu$eV. In 
such conditions the minimization becomes almost impossible to converge, the MLWFs cannot be calculated for SO-split 
bands and an alternative scheme must be implemented.

Here we describe a method for obtaining the SO matrix elements with respect to the Wannier functions calculated in 
the absence of the SO interaction. Since the SO coupling in OSCs is weak, such spin-independent Wannier functions 
represent a close approximation of those that one could, at least in principle, obtain in the presence of the SO 
interaction. Furthermore, when the MLWF basis spans the same Hilbert space defined by all the atomic orbitals
relevant for describing a given bands manifold, our method provides an accurate description of the system even in the
case of heavy elements, i.e. for strong spin-orbit interaction. In particular we implement our scheme together with the
atomic-orbital, pseudopotential, density functional theory (DFT) code {\sc Siesta}~\cite{0953-8984-14-11-302}. {\sc Siesta} is
used to generate the band structure in absence of the spin-orbit interaction and for calculating the SO potential, while
the MLWF procedure is performed with the {\sc Wannier90} code~\cite{Mostofi2008685}. 

The paper is organized as follows. In the next section we describe our method in detail, by starting from the general idea
and then going into the specific numerical implementation. A how-to workflow will also be presented. Next we discuss
results obtained for rather diverse physical systems. Firstly, we evaluate the SO-split energy eigenvalues of a plumbane 
molecule and show how accurately these match those obtained directly from DFT including SO interaction. Then, we apply 
our procedure to the calculation of the band structure of a chain of Pb atoms, before moving to materials composed of light 
elements with low SO coupling. Here we will show that our method performs well for chains made of carbon atoms and of 
methane molecules. Finally we obtain the SO matrix elements for the Wannier functions derived from the HOMO band of a 
triarylamine-based nanowire, a relatively well-known semiconducting material with potential applications in 
photo-voltaic~\cite{B908802D} and spintronics.

\section{Method}

\subsection{General idea}

Here we describe the idea behind our method, which is general and does not depend on the specific implementation
used for calculating the band structure. Consider a set of $N^\prime$ isolated Bloch states, $\ket{\psi_{m\mathbf{k}}}$, 
describing an infinite lattice. These for instance can be the DFT Kohn-Sham eigenstates of a crystal. One can then obtain 
the associated $N^\prime$ Wannier functions from the definition,
\begin{equation}\label{equ1}
\begin{split}
\ket{w_{n\mathbf{R}}}=\frac{V}{(2\pi)^3}\int_\mathrm{BZ}\left[\sum_{m=1}^{N^\prime} U_{mn}^{\mathbf{k}} 
\ket{\psi_{m\mathbf{k}}}\right]e^{-i\mathbf{k}\cdot\mathbf{R}}d\mathbf{k}\:,
\end{split}
\end{equation}
where $\ket{w_{n\mathbf{R}}}$ is the $n$-th Wannier vector centred at the lattice site $\mathbf{R}$, $V$ is the volume of 
the primitive cell and the integration is performed over the first Brillouin zone (BZ). In Eq.~(\ref{equ1}) $U^{\mathbf{k}}$ is 
a unitary operator that mixes the Bloch states and hence defines the specific set of Wannier functions. A particularly convenient
gauge choice for $U^{\mathbf{k}}$ consists in minimizing the Wannier functions spread, which writes
\begin{equation}\label{equ2}
\begin{split}
\Omega=\sum_n\left[\bra{w_{n\mathbf{0}}}r^2\ket{w_{n\mathbf{0}}}-|\bra{w_{n\mathbf{0}}}\mathbf{r}\ket{w_{n\mathbf{0}}}|^2\right]\:.
\end{split}
\end{equation}
Such choice defines the so-called maximally localized Wannier functions (MLWFs). 

In the absence of SO coupling a Wannier function of spin $s_1$ is composed exclusively of Bloch states 
with the same spin, $s_1$. By moving from a continuos to a discrete $k$-point representation the spin-polarized 
version of Eq.~(\ref{equ1}) becomes \cite{RevModPhys.84.1419}
\begin{equation}\label{equ9}
\begin{split}
\ket{w^{s_1}_{n\mathbf{R}}}=\frac{1}{N}\sum_{\mathbf{k}}\sum_m U^{s_1}_{mn}(\mathbf{k}) 
\ket{\psi_{m\mathbf{k}}^{s_1}}e^{-i\mathbf{k}\cdot\mathbf{R}}\:.
\end{split}
\end{equation}
Note that this represents either a finite periodic lattice comprising $N$ unit cells or a sampling of $N$ uniformly 
distributed $k$-points in the Brillouin zone of an infinite lattice. Here the Bloch states, which are normalized 
within each unit cell according to the relation 
$\braket{\psi_{m\mathbf{k}}^{s_1}|\psi_{n\mathbf{k'}}^{s_2}}=N\delta_{m,n}\delta_{\mathbf{k},\mathbf{k'}}\delta_{s_1,s_2}$, 
obey to the condition $\psi_{p\mathbf{k}}(\mathbf{r}_1)=\psi_{p\mathbf{k}}(\mathbf{r}_{N+1})$, where 
$\psi_{p\mathbf{k}}(\mathbf{r}_m)$ denotes the Bloch function for the $p$-th band at the wavevector \textbf{k} 
and position $\mathbf{r}_m$. 

The projection of a generic Bloch state onto a MLWF in the absence of SO coupling can be written as 
\begin{multline}\label{equ10}
\begin{split}
&\braket{{\psi_{q\mathbf{k'}}^{s_1}}|{w_{n\mathbf{R}_2}^{s_2}}}= \\
&=\frac{1}{N}\sum_{\mathbf{k}}\sum_m U_{mn}^{s_2} (\mathbf{k})\braket{{\psi_{q\mathbf{k'}}^{s_1}}|{\psi_{m\mathbf{k}}^{s_2}}}
e^{-i\mathbf{k}\cdot\mathbf{R}}=\\
&=\frac{1}{N}\sum_{\mathbf{k}}\sum_m U^{s_2}_{mn}(\mathbf{k})e^{-i\mathbf{k}.\mathbf{R}}N\delta_{q,m}
\delta_{\mathbf{k},\mathbf{k'}}\delta_{s_1,s_2}=\\
&=U^{s_2}_{qn}(\mathbf{k'})e^{-i\mathbf{k'}\cdot\mathbf{R}}\delta_{s_1,s_2}\:.
\end{split}
\end{multline}
Hence a generic SO matrix element can be expanded over the MLWF basis set as
\begin{multline}\label{equ11}
\begin{split}
&\bra{w_{m\mathbf{R}_1}^{s_1}}\mathbf{V}_\mathrm{SO}\ket{w_{n\mathbf{R}_2}^{s_2}}= \\
&=\frac{1}{N^2}\sum_{p,q}\sum_{\mathbf{k}_1,\mathbf{k}_2}\braket{{w_{m\mathbf{R}_1}^{s_1}}|{\psi_{p\mathbf{k}_1}^{s_1}}}
(\mathbf{V}_\mathrm{SO})^{s_1,s_2}_{p\mathbf{k}_1,q\mathbf{k}_2}
%\bra{\psi_{p\mathbf{k}_1}^{s_1}}\mathbf{V}_\mathrm{SO}\ket{\psi_{q\mathbf{k}_2}^{s_2}}
\braket{\psi_{q\mathbf{k}_2}^{s_2}|{w_{n\mathbf{R}_2}^{s_2}}}=\\
&=\frac{1}{N^2}\sum_{p,q}\sum_{\mathbf{k}_1,\mathbf{k}_2}U^{*(s_1)}_{pm}(\mathbf{k}_1)e^{i\mathbf{k}_1\cdot\mathbf{R}_1}
(\mathbf{V}_\mathrm{SO})^{s_1,s_2}_{p\mathbf{k}_1,q\mathbf{k}_2}\cdot\\
& \quad \cdot U^{s_2}_{qn}(\mathbf{k}_2)e^{-i\mathbf{k}_2\cdot\mathbf{R}_2}\:,
%& \quad \bra{\psi_{p\mathbf{k}_1}^{s_1}}\mathbf{V}_{SO}\ket{\psi_{q\mathbf{k}_2}^{s_2}}\:,
\end{split}
\end{multline}
where
\begin{equation}\label{VSOME}
(\mathbf{V}_\mathrm{SO})^{s_1,s_2}_{p\mathbf{k}_1,q\mathbf{k}_2}=
\bra{\psi_{p\mathbf{k}_1}^{s_1}}\mathbf{V}_\mathrm{SO}\ket{\psi_{q\mathbf{k}_2}^{s_2}}\:.
\end{equation}

It must be noted that in the absence of SO coupling, the Bloch states are spin-degenerate, i.e. there are two states 
corresponding to each spatial wave-function, one with spin up, 
$\ket{\psi^{\uparrow}(\mathbf{r})}=\ket{\psi(\mathbf{r})}\otimes\ket{\uparrow}$, and one with spin down, 
$\ket{\psi^{\downarrow}(\mathbf{r})}=\ket{\psi(\mathbf{r})}\otimes\ket{\downarrow}$. The same is true for the 
Wannier functions, i.e. one has always the pair $\ket{w^{\uparrow}(\mathbf{r})}=\ket{w(\mathbf{r})}\otimes\ket{\uparrow}$,
$\ket{w^{\downarrow}(\mathbf{r})}=\ket{w(\mathbf{r})}\otimes\ket{\downarrow}$.
In the presence of SO coupling, spin mixing occurs and each Bloch and Wannier state is, in general, a linear combination
of both spin vectors. Since the Bloch states (or the Wannier ones) obtained in the absence of SO coupling form a 
complete basis set in the Hilbert space, the SO coupling operator can be written over such basis provided that one takes 
both spins into account. Therefore we use such spin-degenerate states as our basis for all calculations.

\subsection{Numerical Implementation}
The derivation leading to Eq.~(\ref{equ11}) is general and the final result is simply a matrix transformation of the SO 
operator from the basis of the Bloch states to that of Wannier ones. Note that both basis sets are those calculated 
in the absence of SO coupling, i.e. we have assumed that the spatial part of the basis function is not modified by the 
introduction of the SO interaction. For practical purposes we now we wish to re-write Eq.~(\ref{equ11}) in terms of a 
localized atomic-orbital basis set, i.e. we wish to make our method applicable to first-principles DFT calculations 
implemented over local orbitals. In particular all the calculations that will follow use the {\sc Siesta} package, which
expands the wave-function and all the operators over a numerical atomic-orbital basis sets,
\{${\ket{\phi_{\mu,\mathbf{R}_j}^{s}}}$\}, where $\ket{\phi_{\mu,\mathbf{R}_j}^{s}}$ denotes the $\rm{\mu}$-th 
atomic orbital ($\mu$ is a collective label for the principal and angular momentum quantum numbers) with spin 
$s$ belonging to the cell at the position $\mathbf{R}_j$. {\sc Siesta} uses relativistic pseudopotentials to generate 
the spin-orbit matrix elements with respect to the basis vectors and truncates the range of the SO interaction to 
the on-site terms~\cite{0953-8984-18-34-012}. For a finite periodic lattice comprising $N$ unit cells, a Bloch state 
is written with respect to atomic orbitals as 
\begin{equation}\label{equ3}
\begin{split}
\ket{\psi_{p\mathbf{k}}}=\sum_{j=1}^N e^{i\mathbf{k}\cdot\mathbf{R}_j}
\left(\sum_{\mu}C_{\mu p}(\mathbf{k})\ket{\phi_{\mu,\mathbf{R}_j}}\right)\:,
\end{split}
\end{equation}
where the coefficients $C_{\mu p}(\mathbf{k})$ are in general C-numbers. This state is normalized over unit cell 
with the allowed $\rm{\mathbf{k}}$-values being $\frac{m}{N} \mathbf{K}$, where $\rm{\mathbf{K}}$ is the reciprocal 
lattice vector and $m$ is an integer. 

%It is worth noting that although this normalization convention differs from that of the Bloch states used in Eq.~(\ref{equ9}) or in SIESTA, we stick to this convention for notational simplicity. This does not alter the final results in any way.

Hence the SO matrix elements written with respect to the spin-degenerate Bloch states calculated in absence of 
SO interaction are
\begin{multline}\label{equ4}
\begin{split}
&\bra{\psi_{p\mathbf{k}_1}^{s_1}}\mathbf{V}_\mathrm{SO}\ket{\psi_{q\mathbf{k}_2}^{s_2}} =
\sum_{j,l}e^{i(\mathbf{k}_2\cdot\mathbf{R}_{l}-\mathbf{k}_1\cdot\mathbf{R}_{j})}\cdot\\
& \cdot\sum_{\mu,\nu}C_{\mu p}^{*s_1}(\mathbf{k}_1)C_{\nu q}^{s_2}(\mathbf{k}_2)
\bra{\phi_{\mu,\mathbf{R}_{j}}^{s_1}}\mathbf{V}_\mathrm{SO}\ket{\phi_{\nu,\mathbf{R}_{l}}^{s_2}}\:.
\end{split}
\end{multline}
As mentioned above {\sc Siesta} neglects all the SO matrix elements between atomic orbitals located at different atoms. This 
leads to the approximation 
\begin{multline}\label{equ5}
\begin{split}
\bra{\phi_{\mu,\mathbf{R}_{j}}^{s_1}}\mathbf{V}_\mathrm{SO}\ket{\phi_{\nu,\mathbf{R}_{l}}^{s_2}}=
\bra{\phi_{\mu}^{s_1}}\mathbf{V}_\mathrm{SO}\ket{\phi_{\nu}^{s_2}}\delta_{\mathbf{R}_{j},\mathbf{R}_{l}}\:,
\end{split}
\end{multline}
so that Eq.~(\ref{equ4}) becomes
\begin{multline}\label{equ6}
\begin{split}
&\bra{\psi_{p\mathbf{k}_1}^{s_1}}\mathbf{V}_\mathrm{SO}\ket{\psi_{q\mathbf{k}_2}^{s_2}} =
\sum_je^{i(\mathbf{k}_2-\mathbf{k}_1)\cdot\mathbf{R}_j}\cdot\\
& \cdot\sum_{\mu,\nu}C_{\mu p}^{*(s_1)}(\mathbf{k}_1)C_{\nu q}^{(s_2)}(\mathbf{k}_2)\bra{\phi_{\mu}^{s_1}}
\mathbf{V}_\mathrm{SO}\ket{\phi_{\nu}^{s_2}}\:.
\end{split}
\end{multline}
This can be further simplified by taking into account the relation
\begin{equation}\label{equ7}
\begin{split}
\sum_{j=1}^N e^{i(\mathbf{k}_1-\mathbf{k}_2)\cdot\mathbf{R}_j}=N\delta_{\mathbf{k}_1,\mathbf{k}_2}\:,
\end{split}
\end{equation}
which leads to the final expression for the SO matrix elements
\begin{multline}\label{equ8}
\begin{split}
&\bra{\psi_{p\mathbf{k}_1}^{s_1}}\mathbf{V}_\mathrm{SO}\ket{\psi_{q\mathbf{k}_2}^{s_2}} =\\
&=N\sum_{\mu,\nu}C_{\mu p}^{*(s_1)}(\mathbf{k}_1)C_{\nu q}^{(s_2)}(\mathbf{k}_1) 
\bra{\phi_{\mu}^{s_1}}\mathbf{V}_\mathrm{SO}\ket{\phi_{\nu}^{s_2}}\delta_{\mathbf{k}_1,\mathbf{k}_2}\:.
\end{split}
\end{multline}

With the result of Eq.~(\ref{equ8}) at hand we can now come back to the expression for the SO matrix elements 
written over the MLWFs computed in absence of spin-orbit [see Eq.~(\ref{equ11})]. In the case of the {\sc Siesta} 
basis set this now reads
\begin{multline}\label{equ12}
\begin{split}
&\bra{w_{m\mathbf{R}_1}^{s_1}}\mathbf{V}_\mathrm{SO}\ket{w_{n\mathbf{R}_2}^{s_2}}= \\
=&\frac{1}{N}\sum_{p,q,\mu,\nu}\sum_{\mathbf{k}}C_{\mu p}^{*s_1}(\mathbf{k})C_{\nu q}^{s_2}(\mathbf{k})
U^{*(s_1)}_{pm}(\mathbf{k})U^{s_2}_{qn}(\mathbf{k})\cdot\\
&\cdot e^{i\mathbf{k}\cdot(\mathbf{R}_1-\mathbf{R}_2)}\bra{\phi_{\mu}^{s_1}}\mathbf{V}_\mathrm{SO}\ket{\phi_{\nu}^{s_2}}\:.
\end{split}
\end{multline}
Finally, we go back to the continuous representation ($N\rightarrow\infty$), where the sum over \textbf{k} is replaced 
by an integral over the first Brillouin zone
\begin{multline}\label{equ13}
\begin{split}
&\bra{w_{m\mathbf{R}_1}^{s_1}}\mathbf{V}_\mathrm{SO}\ket{w_{n\mathbf{R}_2}^{s_2}}= \\
&=\frac{V}{(2\pi)^3}\sum_{p,q,\mu,\nu}\int_\mathrm{BZ}C_{\mu p}^{*s_1}(\mathbf{k})C_{\nu q}^{s_2}(\mathbf{k})
U^{*s_1}_{pm}(\mathbf{k})U^{s_2}_{qn}(\mathbf{k})\cdot\\
&\cdot e^{i\mathbf{k}\cdot(\mathbf{R}_1-\mathbf{R}_2)}\bra{\phi_{\mu}^{s_1}}
\mathbf{V}_\mathrm{SO}\ket{\phi_{\nu}^{s_2}}d\mathbf{k}\:.
\end{split}
\end{multline}

To summarize, our strategy consists in simply evaluating the SO matrix elements over the basis set of the MLWFs
constructed in the absence of SO interaction. These are by definition spin-degenerate and they are in general easy
to compute since associated to well-separated bands. Our procedure thus avoids to run the minimization algorithm
necessary to fix the Wannier's gauge over the SO-split bands, which in the case of OSCs have tiny splits. Our method
is exact in the case the MLWFs form a complete set describing a particular bands manifold. In other circumstances 
they constitute a good approximation, as long as the SO interaction is weak, namely when it does not change significantly
the spatial shape of the Wannier functions. However, for a material with strong SO coupling (eg. Pb), if the MLWFs under 
consideration do not span the entire Bloch states manifold, then the SO-split eigenvalues calculated with our method 
will not match those obtained directly with the first principles calculation.

\subsection{Workflow}

The following procedure is adopted when calculating the SO-split band structures from the MLWFs Hamiltonian. 
The results are then compared to the band structure obtained directly from {\sc Siesta} including SO interaction.

\begin{enumerate}

\item We first run a self-consistent non-collinear spin-DFT {\sc Siesta} calculation and obtain the band structure. 

\item From the density matrix obtained at step (1), we run a non self-consistent single-step {\sc Siesta} calculation 
including SO coupling. This gives us the matrix elements 
$\bra{\phi_{\mu}^{s_1}}\mathbf{V}_\mathrm{SO}\ket{\phi_{\nu}^{s_2}}$. 
The band structure obtained in this calculation (from now on this is called the SO-DFT band structure) will 
be then compared with that obtained over the MLWFs. 
Note that we do not perform the {\sc Siesta} DFT calculation including spin-orbit interaction in a self-consistent 
way. This is because the SO interaction changes little the density matrix so that such calculation is often not
necessary. Furthermore, as we cannot run the MLWF calculation in a self-consistent way over the SO 
interaction, considering non-self-consistent SO band structure at the {\sc Siesta} level allows us to compare
electronic structures arising from identical charge densities. 

\item Since the current version of {\sc Wannier90} implemented for {\sc Siesta} works only with collinear spins, we run 
a regular self-consistent spin-polarized {\sc Siesta} calculation. This gives us the coefficients $C_{\mu n}^{s}(\mathbf{k})$, 
which are spin-degenerate for a non-magnetic material, 
$C_{\mu n}^{\uparrow}(\mathbf{k})=C_{\mu n}^{\downarrow}(\mathbf{k})$. 

\item We run a {\sc Wannier90} calculation to construct the MLWFs associated to the band structure computed at
point (3). This returns us the unitary matrix, $U_{pm}^{s}(\mathbf{k})$, the Hamiltonian matrix elements 
$\bra{w_{m\mathbf{R}_1}^{s_1}}\mathbf{H}_0\ket{w_{n\mathbf{R}_2}^{s_2}}$ ($\mathbf{H}_0$ is the Kohn-Sham
Hamiltonian in absence of SO interaction) and the phase 
factors~\footnote{The correctness of the  elements $U_{pm}^{s}(\mathbf{k})$ and $e^{i\mathbf{k}\cdot\mathbf{R}}$ 
is easily verified by ensuring that the following relation is satisfied
\begin{multline}\label{equ_footnote}
\begin{split}
 \braket{w_{m\mathbf{R}_1}|w_{n\mathbf{R}_2}}&=\frac{1}{N}\sum_p \int_\mathrm{FBZ} 
 d\mathbf{k} \braket{w_{m\mathbf{R}_1}|\psi_{p\mathbf{k}}}\braket{\psi_{p\mathbf{k}}|w_{n\mathbf{R}_2}}=\\ 
 &=\frac{1}{N}\sum_p \int_\mathrm{FBZ} d\mathbf{k}U^*_{pm}(\mathbf{k})U_{pn}(\mathbf{k})e^{i\mathbf{k}\cdot
 (\mathbf{R}_1-\mathbf{R}_2)}=\\ 
 &=\delta_{m,n}\delta_{\mathbf{R}_1,\mathbf{R}_2}  \:.
\end{split}
\end{multline}
} 
$e^{i\mathbf{k}\cdot\mathbf{R}}$. For a non-magnetic material the matrix elements of $\mathbf{H}_0$ satisfy
the relation $\bra{w_{m\mathbf{R_1}}^{s_1}}\mathbf{H}_0\ket{w_{n\mathbf{R_2}}^{s_2}}=
 \bra{w_{m\mathbf{R_1}}}\mathbf{H}_0\ket{w_{n\mathbf{R_2}}}\delta_{s_1,s_2}$.

\item From $\bra{\phi_{\mu}^{s_1}}\mathbf{V}_\mathrm{SO}\ket{\phi_{\nu}^{s_2}}$ and the
$C_{\mu n}^{s}(\mathbf{k})$'s we calculate the matrix elements 
$\bra{\psi_{p\mathbf{k}}^{s_1}}\mathbf{V}_\mathrm{SO}\ket{\psi_{q\mathbf{k}}^{s_2}}$ 
by using Eq.~(\ref{equ8}).

\item Next we transform the SO matrix elements constructed over the Bloch functions, 
$\bra{\psi_{p\mathbf{k}}^{s_1}}\mathbf{V}_\mathrm{SO}\ket{\psi_{q\mathbf{k}}^{s_2}}$,
into their Wannier counterparts, 
$\bra{w_{m\mathbf{R}_1}^{s_1}}\mathbf{V}_\mathrm{SO}\ket{w_{n\mathbf{R}_2}^{s_2}}$,
by using Eq.~(\ref{equ13}).

\item The final complete Wannier Hamiltonian now reads
\begin{multline}\label{equ14}
\begin{split}
 \bra{w_{m\mathbf{R}_1}^{s_1}}\mathbf{H}\ket{w_{n\mathbf{R}_2}^{s_2}}
 &=\bra{w_{m\mathbf{R}_1}^{s_1}}\mathbf{H}_0+\mathbf{V}_\mathrm{SO}\ket{w_{n\mathbf{R}_2}^{s_2}}\:,
\end{split}
\end{multline}
and the associated band structure can be directly compared with that computed at point (2) directly
from {\sc Siesta}.

\end{enumerate}

\section{Results and Discussion}

We now present our results, which are discussed in the light of the theory just described. 

\subsection{Plumbane Molecule}

\begin{figure}
\centering
 \includegraphics[width=0.44\textwidth]{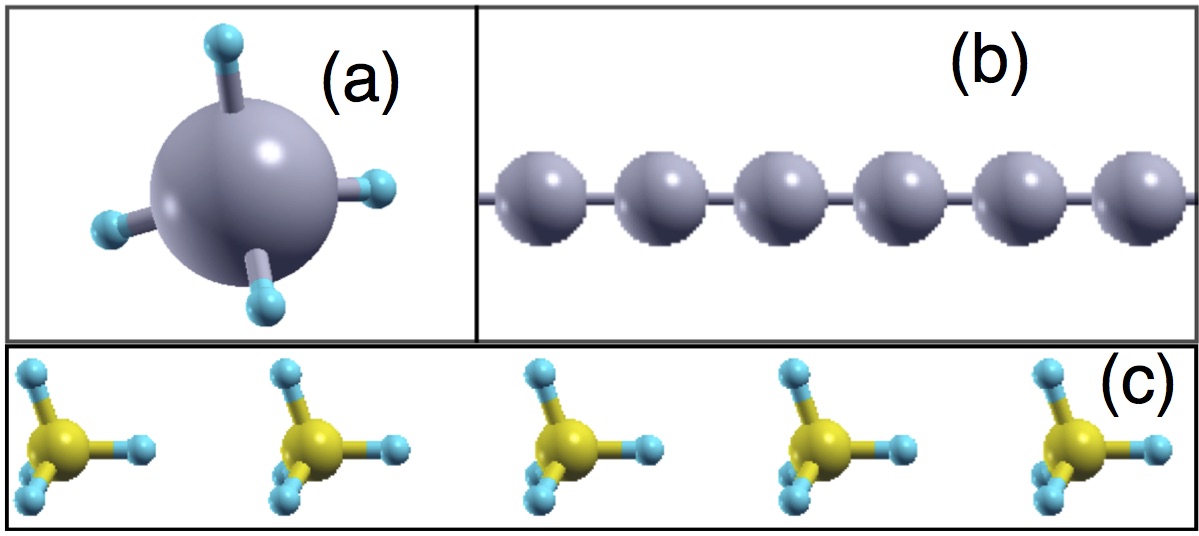}%
\caption{(Color on line) Atomic structure of (a) a plumbane molecule, (b) a chain of lead atoms and (c) a chain of 
methane molecules. We have also calculated the electronic structure of a chain of C atoms, which is essentially 
identical to that presented in (b). Color code: Pb = grey, H = light blue, C = yellow.}\label{fig:three_structure}
\label{fig:3_structures}
\end{figure}

We start our analysis by calculating the SO matrix elements and then the energy eigenvalues of a plumbane, PbH$_4$, 
molecule [see figure~\ref{fig:3_structures}(a)]. 
Due to the presence of lead, the molecular eigenstates change significantly when the SO interaction is switched on. 
For this non-periodic system the key relations in Eq.~(\ref{equ8}) and Eq.~(\ref{equ11}) reduce to 
\begin{multline}\label{equ16}
\begin{split}
\bra{\psi_{p}^{s_1}}\mathbf{V}_\mathrm{SO}\ket{\psi_{q}^{s_2}} =
\sum_{\mu,\nu}C_{\mu p}^{*s_1}C_{\nu q}^{s_2}\bra{\phi_{\mu}^{s_1}}
\mathbf{V}_\mathrm{SO}\ket{\phi_{\nu}^{s_2}}
\end{split}
\end{multline}
and 
\begin{multline}\label{equ17}
\begin{split}
&\bra{w_{m}^{s_1}}\mathbf{V}_\mathrm{SO}\ket{w_{n}^{s_2}} =
\sum_{p,q}U^{*s_1}_{pm}U^{s_2}_{qn}\bra{\psi_{p}^{s_1}}
\mathbf{V}_\mathrm{SO}\ket{\psi_{q}^{s_2}}\:,
\end{split}
\end{multline}
respectively, where now the vectors $\psi_n^s$ are simply the eigenvectors with quantum number $n$ and
spin $s$.

In Table~\ref{table:plumbane} we report the first 10 energy eigenvalues of plumbane, calculated either with or 
without SO coupling. These have been computed within the LDA (local density approximation) and a double-zeta 
polarized basis set. The table compares results obtained with our MLWFs procedure to those computed with 
SO-DFT by {\sc Siesta}. Clearly in this case of a heavy ion the SO coupling changes the eigenvalues appreciably,
in particular in the spectral region around -13~eV. Such change is well captured by our Wannier calculation, which
returns energy levels in close proximity to those computed with SO-DFT by {\sc Siesta}. In order to estimate the error 
introduced by our method, we calculate the \textit{Mean Relative Absolute Difference (MRAD)}, which we define 
as $\frac{1}{N}\sum\frac{|\epsilon_i^s-\epsilon_i^w|}{|\epsilon_i^s|}$ for a set of $N$ eigenvalues ($i=1,...,N$), 
where $\epsilon_i^s$ and $\epsilon_i^w$ are the $i$-th eigenvalues calculated from {\sc Siesta} and the MLWFs, 
respectively. Notably the \textit{MRAD} is rather small both in the SO-free case and when the SO interaction
is included. Most importantly, we can report that our procedure to evaluate the SO matrix elements over the MLWFs 
basis clearly does not introduce any additional error.
\begin{table}
\centering
\begin{tabular}{>{\centering}p{2cm}>{\centering}p{2cm}| >{\centering}p{2cm}  cp{2cm} }
\hline  \multicolumn{2}{c}{NonSO}  & \multicolumn{2}{c}{SO} \\ \hline
{\sc Siesta}           & MLWF         & {\sc Siesta}                             & MLWF          \\ \hline
 -33.93534                            &-33.93521   &-33.93532        &-33.93521       \\
 -33.93530                            &-33.93521    &-33.93528      &-33.93521       \\
 -13.02511                            &-13.02507    &-14.69573       &-14.69568      \\
 -13.02511                            &-13.02507     &-14.69573      &-14.69568      \\
 -13.02510                            &-13.02506    &-12.64301       &-12.64298       \\
 -13.02509                            &-13.02506     &-12.64301      &-12.64298       \\
 -13.02320                            &-13.02315    &-12.64166       &-12.64162       \\
 -13.02318                            &-13.02315    &-12.64165        &-12.64162       \\
 -5.75256                              &-5.75251       &-5.75255      &-5.75251       \\
 -5.75245                                &-5.75251     &-5.75245       &-5.75251  \\ \hline
\multicolumn{2}{c}{\textit{MRAD}=$4.320\times10^{-6}$}  & \multicolumn{2}{c}{\textit{MRAD}=$3.998\times10^{-6}$} \\ \hline
 \end{tabular}
\captionsetup{justification=justified, singlelinecheck=false}
     \caption{The 10 lowest energy eigenvalues of a plumbane molecule calculated with (SO) and without (NonSO) 
     spin-orbit interaction. The first and third columns correspond to the SO-DFT {\sc Siesta} calculation, 
     while the second and the fourth to the MLWFs one. The \textit{MRAD} for both cases is reported in the last row.}
     \label{table:plumbane}
\end{table}

Before discussing some of the properties of the SO matrix elements associated to this particular case of 
a finite molecule, we wish to make a quick remark on the Wannier procedure adopted here. The eigenvalues 
reported in Table~\ref{table:plumbane} are the ten with the lowest energies. However, in order to construct the 
MLWFs we have considered all the states of the calculated Kohn-Sham spectrum. This means that, if our 
{\sc Siesta} basis set describes PbH$_4$ with $N$ distinct atomic orbitals, then the MLWFs constructed
are 2$N$ (the factor 2 accounts for the spin degeneracy). In this case the original local orbital basis set and the 
constructed MLWFs span the same Hilbert space and the mapping is exact, whether or not the SO interaction is
considered. 

In most cases, however, one wants to construct the MLWFs by using only a subset of the spectrum, for instance
the first $N^\prime$ eigenstates. Since in general the SO interaction mixes all states, there will be SO matrix
elements between the selected $N^\prime$ states and the remaining $N-N^\prime$. This means that a MLWF 
basis constructed only from the first $N^\prime$ eigenstates will not be able to provide an accurate description of 
the SO-split spectrum. Importantly, one in general may expect that the SO interaction matrix elements between
different Kohn-Sham orbitals, $\bra{\psi^{s_1}_p}\mathbf{V}_{\rm{SO}}\ket{\psi^{s_2}_q}$, are smaller than those
calculated at the same orbital, $\bra{\psi^{s_1}_n}\mathbf{V}_{\rm{SO}}\ket{\psi^{s_2}_n}$. This is because of
the short-range of the SO interaction and the fact that the Kohn-Sham eigenstates are orthonormal. In the case
of light elements, i.e. for a weak SO potential, one may completely neglect the off-diagonal SO matrix elements.
This means that the SO spectrum constructed with the MLWFs associated to the first $N^\prime$ eigenstates
will be approximately equal to the first $N^\prime$ eigenvalues of the MLWFs Hamiltonian constructed over the 
entire $N$-dimensional spectrum. Such property is particularly relevant for OSCs, for which the SO interaction is
weak.

We now move to discuss a general property of the MLWF SO matrix elements, namely the relations  
$\bra{w_m^{s}}\mathbf{V}_\mathrm{SO}\ket{w_m^{s}}=0$ and $\Re[\bra{w_m^{s}}\mathbf{V}_\mathrm{SO}\ket{w_n^{s}}]=0$. 
This means that the SO matrix elements for the same spin and the same Wannier function vanish, while those 
for the same spin and different Wannier functions are purely imaginary. This property can be understood from the 
following argument. The SO coupling operator is $\mathbf{V}_\mathrm{SO}=\sum_{\mathbf{R}_j}V_{\mathbf{R}_j}
\mathbf{L}_{\mathbf{R}_j}\cdot\mathbf{S}$, where $V_{\mathbf{R}_j}$ is a scalar potential independent of spin, 
and $\mathbf{L}_{\mathbf{R}_j}$ is the angular momentum operator corresponding to the central potential of the 
atom at position $\mathbf{R}_j$. Here $\mathbf{S}$ is the spin operator and the sum runs over all the atoms.
By now expanding \textbf{S} in terms of the Pauli spin matrices one can see that for any vector 
$\ket{\gamma_i^s}=\ket{\gamma_i}\otimes\ket{s}$, which can be written as a tensor product of a spin-independent 
part, $\ket{\gamma_i}$, and a spinor $\ket{s}$, the following equality holds
\begin{multline}\label{equ20}
\begin{split}
&\bra{\gamma_m^{s_1}}\mathbf{L}\cdot\mathbf{S}\ket{\gamma_n^{s_2}}=
\frac{1}{2}\left[\bra{\gamma_m}\hat{L}_z\ket{\gamma_n}\delta_{s_1\uparrow}\delta_{s_2\uparrow}\right.+\\
&+\bra{\gamma_m}\hat{L}_-\ket{\gamma_n}\delta_{s_1\uparrow}\delta_{s_2\downarrow} +
\bra{\gamma_m}\hat{L}_+\ket{\gamma_n}\delta_{s_1\downarrow}\delta_{s_2\uparrow}+\\
&+\left.\bra{\gamma_m}-\hat{L}_z\ket{\gamma_n}\delta_{s_1\downarrow}\delta_{s_2\downarrow}\right]\:.
\end{split}
\end{multline}
Eq.~(\ref{equ20}) can then be applied to both the Kohn-Sham eigenstates and the MLWFs, since they are both 
written as $\ket{\gamma_i^s}=\ket{\gamma_i}\otimes\ket{s}$.

Now, the atomic orbitals used by {\sc Siesta} have the following form 
\begin{multline}\label{equ21}
\begin{split}
\ket{\phi_i}=\ket{R_{n_i,l_i}}\otimes\ket{l_i,M_i}\:,
\end{split}
\end{multline}
where $\ket{R_{n,l}}$ is a radial numerical function, while the angular dependence is described by the 
real spherial harmonic,~\footnote{The real spherical harmonics are constructed from the complex ones, $\ket{l,m}$, 
as $\ket{l,M}=\frac{1}{\sqrt2}[\ket{l,m}+(-1)^m\ket{l,-m}]$ and $\ket{l,-M}=\frac{1}{i\sqrt2}[\ket{l,m}-(-1)^m\ket{l,-m}]$. 
For $M=0$ the real and complex spherical harmonics coincide.} 
$\ket{l,M}$. It can be proved that the real spherical harmonics follow the relation 
\begin{multline}\label{equ22}
\begin{split}
\bra{l,M_i}\hat{L}_z\ket{l,M_j}=-iM_i\delta_{M_i,M_j}\:.
\end{split}
\end{multline}

Since any Kohn-Sham eigenstate, $\ket{\psi_p^{s_1}}$, can be written as ${\ket{\phi_i}\otimes\ket{s_1}}$, 
Eq.~(\ref{equ20}) implies that only the terms in $\hat{L}_z$ (or $-\hat{L}_z$) contribute to the matrix element
between same spins, $\bra{\psi_p^{s_1}}\mathbf{L}\cdot\mathbf{S}\ket{\psi_p^{s_1}}$. Eq.~(\ref{equ22}) together
with the fact that the Kohn-Sham eigenstates are real for a finite molecule further establishes that 
$\Re[\bra{\psi_p}\hat{L}_z\ket{\psi_q}]=0$. As a consequence $\bra{\psi_m}\hat{L}_z\ket{\psi_m}=0$. 
Finally, by keeping in mind that the unitary matrix elements transforming the Kohn-Sham eigenstates
into MLWFs are real for a molecule, we have also
\begin{multline}
\begin{split}
& \bra{w_m^{s_1}}\mathbf{L}\cdot\mathbf{S}\ket{w_n^{s_1}} =\pm \bra{w_m}\hat{L}_z\ket{w_n}=\\
%&=\sum_{p,q}U_{pm}U_{qn}\bra{\psi_p}\hat{L}_z\ket{\psi_q} =
&=\sum_{p\neq q}U_{pm}U_{qn}\bra{\psi_p}\hat{L}_z\ket{\psi_q}\:,
\end{split}
\end{multline}
which has to be imaginary. Thus we have $\Re\bra{w_m^{s_1}}\mathbf{V}_\mathrm{SO}\ket{w_n^{s_1}}=0$ and 
$\bra{w_m^{s_1}}\mathbf{V}_\mathrm{SO}\ket{w_m^{s_1}}=0$ since $\mathbf{V}_\mathrm{SO}$ must have real 
expectation values.

\subsection{Lead Chain}

 \begin{figure}
\centering{\includegraphics[width=0.41\textwidth]{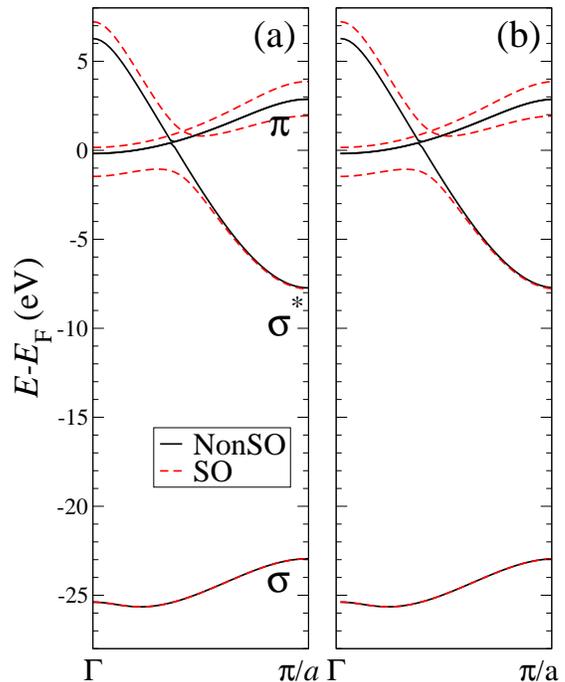}}
\caption{(Color on line) Bandstructure of a 1D Pb chain calculated (a) with {\sc Siesta} and (b) by diagonalizing 
the Hamiltonian matrix constructed over the MLWFs. Black and red lines are for the bands obtained
without and with SO coupling, respectively. The $\sigma$, $\sigma^*$ and $\pi$ bands are identified in
the picture.}\label{fig:figure1}
\label{fig:Pbbands}
\end{figure}

Next we move to calculating the SO matrix elements for a periodic structure. In particular we look at a 1D chain of Pb 
atoms with a unit cell length of 2.55~\AA, which is the DFT equilibrium lattice constant obtained with the LDA. 
Note that free-standing mono-dimensional Pb chains have been never reported in literature, although there are
studies of low-dimensional Pb structures encapsulated into zeolites~\cite{PbZeo}. Here, however, we do not seek
at describing a real compound, but we rather take the 1D Pb mono-atomic chain as a test-bench structure to apply 
our method to a periodic structure with a large SO coupling. Also in this case we have constructed the MLWFs 
by taking the entire bands manifold and not a subset of it. For the DFT calculations we have considered a simple 
$s$ and $p$ single-zeta basis set, which, in absence of SO interaction yields three bands with one of them being 
doubly degenerate [see Fig.~\ref{fig:Pbbands}(a)]. The doubly-degenerate relatively-flat band just cuts across the Fermi 
energy, $E_\mathrm{F}$, and it is composed of the $p_y$ and $p_z$ orbitals orthogonal to the chain axis ($\pi$
band). The other two bands are $sp$ hybrid ($\sigma$ bands). The lowest one at about 25~eV below $E_\mathrm{F}$ 
has mainly $s$ character ($\sigma$ band), while the other mainly $p_x$ ($\sigma^*$ band).

Spin-orbit coupling lifts the degeneracy of the $p$-type band manifold, which is now composed of three distinct 
bands. In particular the degeneracy is lifted only in the $\pi$ band at the edge of the 1D Brillouin zone, while it 
also involves the $\sigma$ one close to the $\Gamma$ point (after the band crossing). When the same band structure 
is calculated from the MLWFs we obtain the plot of Fig.~\ref{fig:Pbbands}(b). This is almost identical to that calculated
with SO-DFT demonstrating the accuracy of our method also for periodic system. 

\begin{figure}
        \centering
\subfloat{\label{fig:real_SOelm}%
  \includegraphics[width=0.39\textwidth]{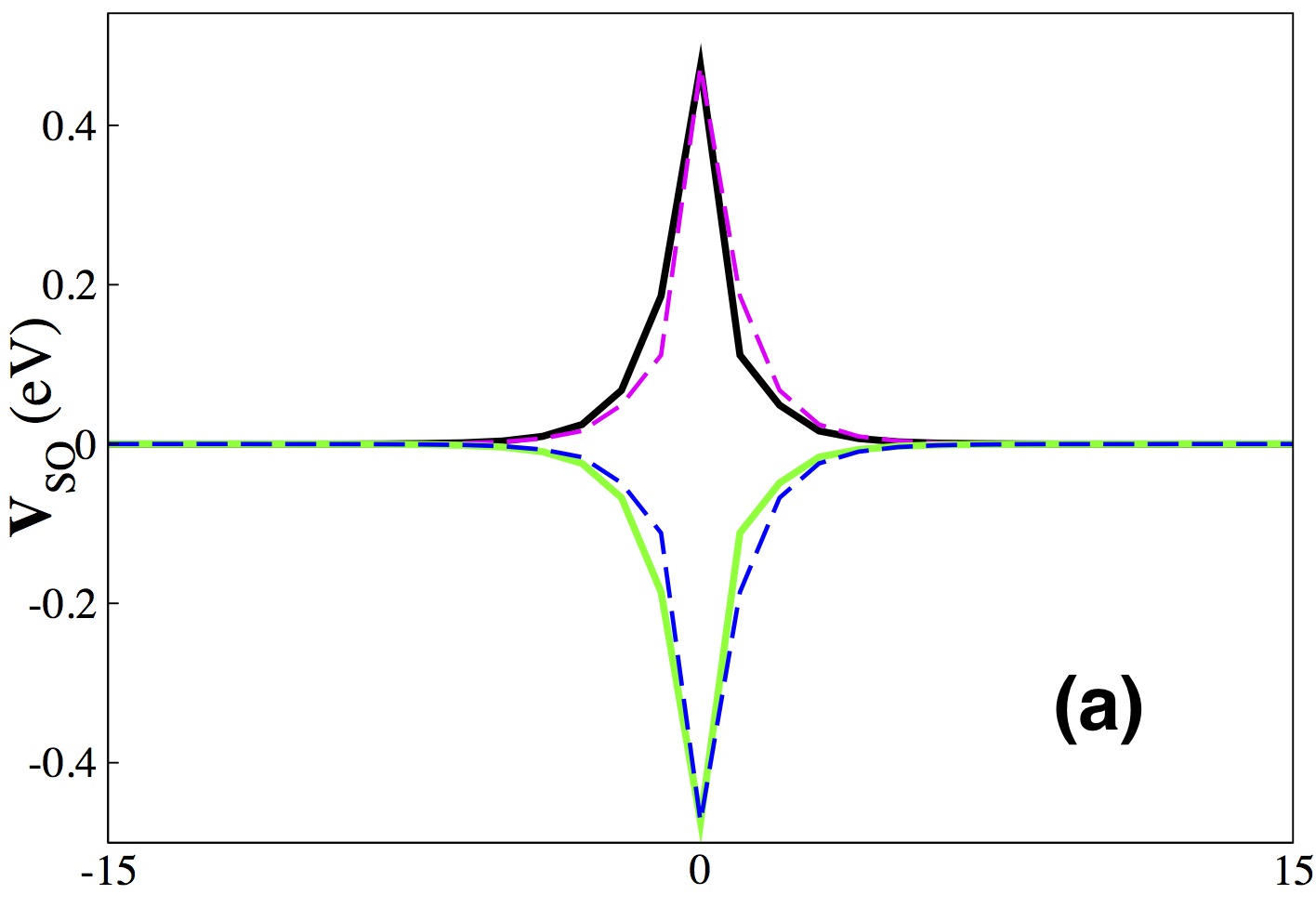}%
}\
\subfloat{\label{fig:aimag_SOelm}%
  \includegraphics[width=0.39\textwidth]{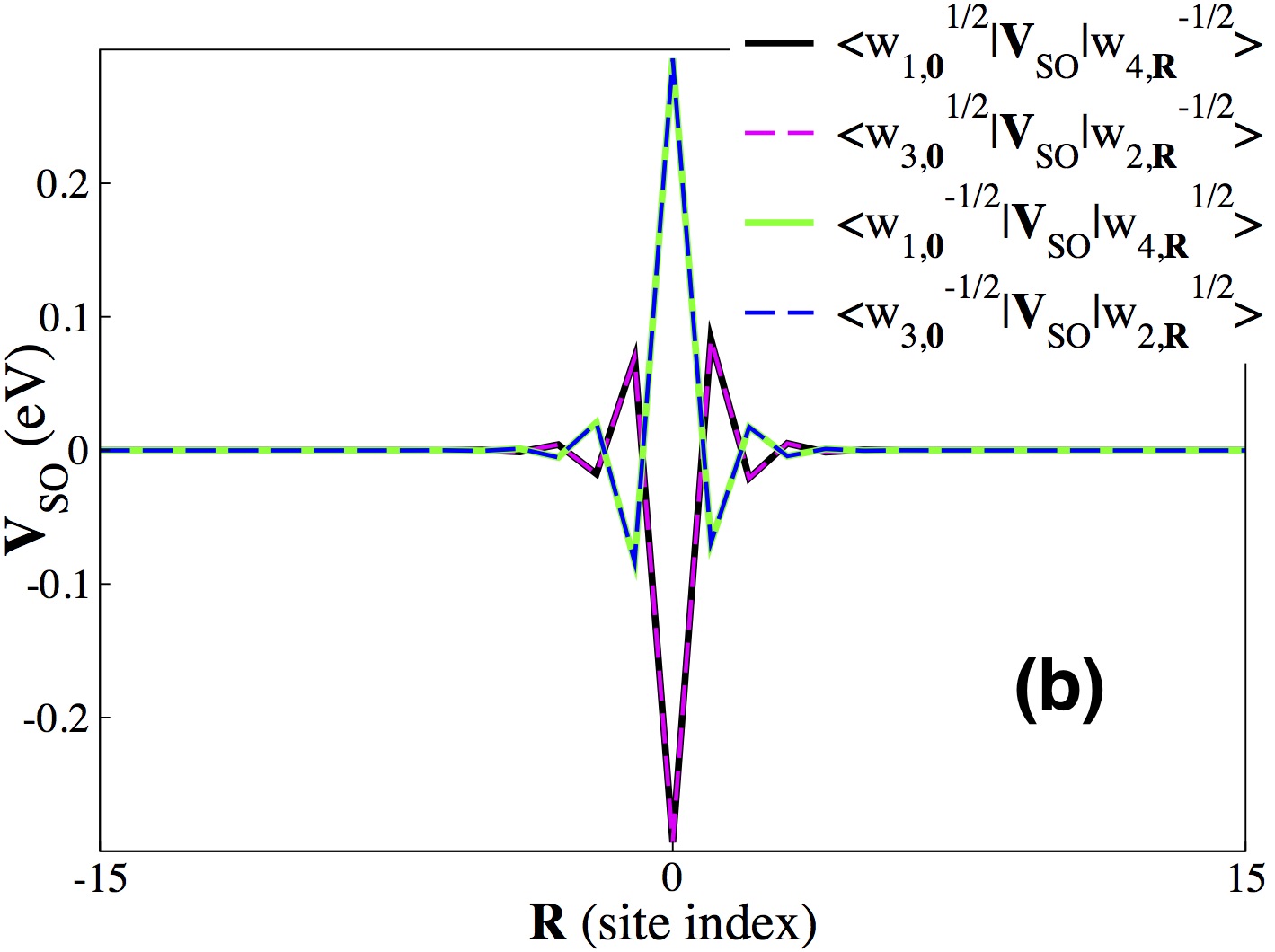}%
}     

\caption{(Color on line) The SO matrix elements of a chain of lead atoms calculated with respect to some representative 
Wannier functions and plotted as a function of the site index, i.e. of the distance between the Wannier function. Panels (a) 
and (b) show the real and imaginary components respectively.}\label{fig:Pb_MATelms}
\end{figure}

It must be noted that for a periodic structure the Bloch state expansion coefficients, $C_{\mu p}(\mathbf{k})$, 
and the elements of the unitary matrix $U$ are complex and consequently the diagonal elements of 
$\mathbf{V}_\mathrm{SO}$ with respect to Wannier functions are not zero in general. However, as expected 
$\bra{w_{m\mathbf{R}}^{s_1}}\mathbf{V}_\mathrm{SO}\ket{w_{n\mathbf{R^\prime}}^{s_2}}$ tends to vanish as 
the separation $|\mathbf{R}-\mathbf{R^\prime}|$ increases. Furthermore, it is clear from Eq.~(\ref{equ20}) that 
the SO matrix elements for Wannier functions should obey the spin-box anti-hermitian relation
\begin{multline}\label{equanti-hermi}
\begin{split}
\bra{w_{m\mathbf{R}}^{s_1}}\mathbf{V}_{SO}\ket{w_{n\mathbf{R'}}^{s_2}}=
-\bra{w_{m\mathbf{R}}^{s_2}}\mathbf{V}_\mathrm{SO}\ket{w_{n\mathbf{R'}}^{s_1}}^{*}\:.
\end{split}
\end{multline}
These two properties can be appreciated in Fig.~\ref{fig:Pb_MATelms}, where we plot the real [panel (a)] and 
imaginary [panel (b)] part of $\bra{w_{m\mathbf{0}}^{s_1}}\mathbf{V}_{SO}\ket{w_{n\mathbf{R}}^{s_2}}$ for some 
representative band combinations, $m$ and $n$, as a function of \textbf{R}.

\subsection{Carbon Chain}

\begin{figure}
\centering{\includegraphics[width=0.48\textwidth]{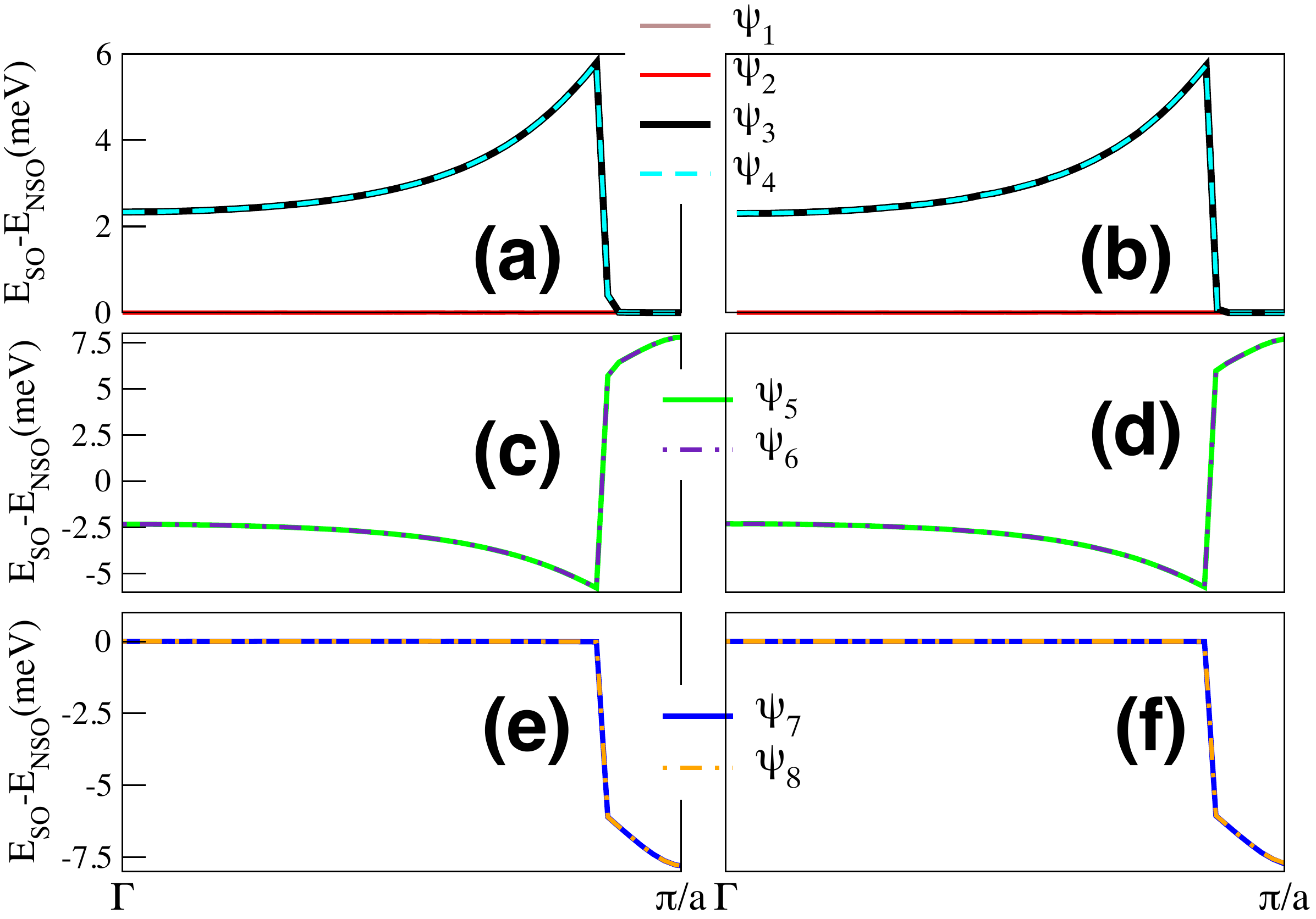}}
\caption{(Color on line) Difference, $E_\mathrm{SO}-E_\mathrm{NSO}$, between the band structure of chain 
of carbon atoms calculated with, $E_\mathrm{SO}$, and without, $E_\mathrm{SO}$, considering SO interaction. 
The bands are labelled in increasing energy order without taking into account spin degeneracy. For instance the 
bands $\psi_1$ and $\psi_2$ are the two spin sub-bands corresponding to the $\sigma$ band (see Fig.~\ref{fig:figure1}
for notation). The left-hand side panels show results for the SO-DFT calculations performed with {\sc Siesta}, 
while the right-hand side one, those obtained from the MLWFs.} \label{fig:Chain_C}
\label{fig:C_BS}
\end{figure}

Next we look at the case of a 1D mono-atomic carbon chain with a LDA-relaxed interatomic distance of 
$\thicksim1.3$~\AA. This has the same structure and electron count of the Pb chain, and the only difference 
concerns the fact that the SO coupling in C is much smaller then that in Pb. In this situation we expect that 
an accurate SO-split band structure can be obtained even when the MLWFs are constructed only for a 
limited number of bands and not for the entire band manifold as in the case of Pb. This time the DFT band 
structure is calculated at the LDA level over a double-zeta polarized (DZP) {\sc Siesta} basis set, comprising
13 atomic orbitals per unit cell. In contrast, the MLWFs are constructed only from the first four bands, which are
well isolated in energy from the rest and again describe the $sp$ bands with $\sigma$ and $\pi$ symmetry. 
Since the SO interaction in carbon is small (the band split is of the order of a few meV) it is impossible to
visualize the effects of the SO interaction in a standard band plot as that in Fig.~\ref{fig:Pbbands}. Hence,
in Fig.~\ref{fig:Chain_C} we plot the difference between the band structure calculated in the presence and in 
the absence of SO coupling. In particular we compare the bands calculated with SO-DFT by {\sc Siesta} 
(left-hand side panels in Fig.~\ref{fig:C_BS}), with those obtained with the MLWFs scheme described here
(right-hand side panels in Fig.~\ref{fig:C_BS}). In the figure we have labelled the bands in order of increasing
energy and neglecting the spin degeneracy. Thus, for instance, the $\psi_1$ and $\psi_2$ bands correspond 
to the two lowest $\sigma$ spin sub-bands (note that the band structure of the linear carbon chain is qualitatively 
identical to that of the Pb one and we can use Fig.~\ref{fig:figure1} to identify the various bands).

We note that the lowest $\sigma$ bands, defined as $\psi_1$ and $\psi_2$, do not split at all due to the
SO interaction, exactly as in the case of Pb. This contrasts the behaviour of both the $\pi$ ($\psi_3$ through 
$\psi_6$) and $\sigma^*$ ($\psi_7$ and $\psi_8$) bands, which instead are modified by the SO interaction.
Notably the changes in energy of the eigenvalues is never larger then 8~meV and it is perfectly reproduced
by our MLWFs representation. This demonstrates that truncating the bands selected for constructing the 
MLWFs is a possible procedure for materials where SO coupling is weak. However, we should note that
the truncation still needs to be carefully chosen. Here for instance we have considered all the 2$s$ and 2$p$ 
bands and neglected those with either higher principal quantum number (e.g. 3$s$ and 3$p$) or higher
angular momentum (e.g. bands with $d$ symmetry originating from the $p$-polarized {\sc Siesta} basis),
which appear at much higher energies. Truncations, where one considers only a particular orbital of a given 
shell (say the $p_z$ orbital in an $np$ shell), need to be carefully assessed since it is unlikely that a clear
energy separation between the bands takes place.
 
 \subsection{Methane Chain}
 
\begin{figure}
\centering
\includegraphics[width=0.50\textwidth]{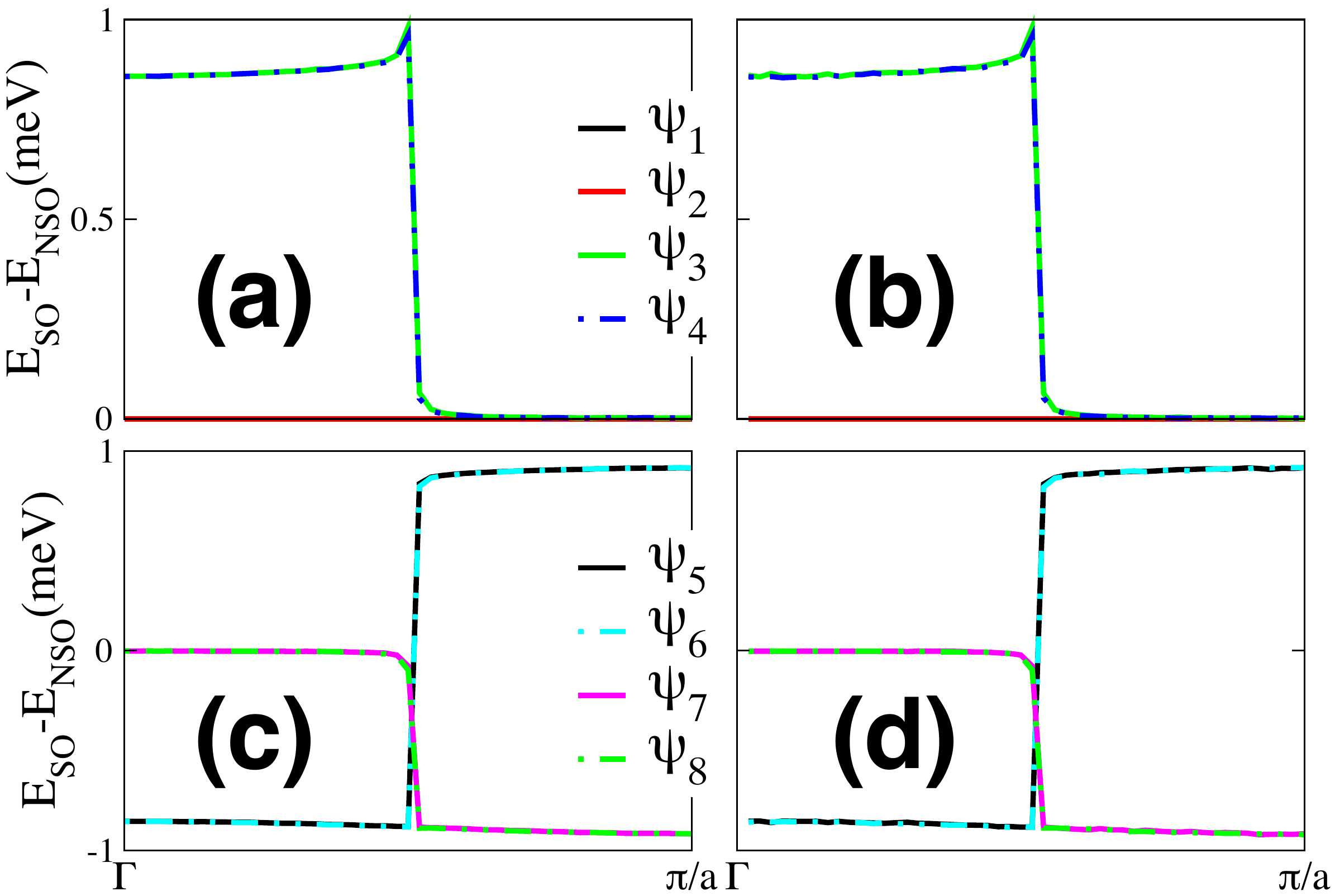}%\llap{\makebox[\wd1]{\raisebox{4cm}{\includegraphics[height=1.6cm]{C_00002_BLACK.png}}}}%
\begin{picture}(0,0)
\put(-51,116){\includegraphics[height=1.8cm]{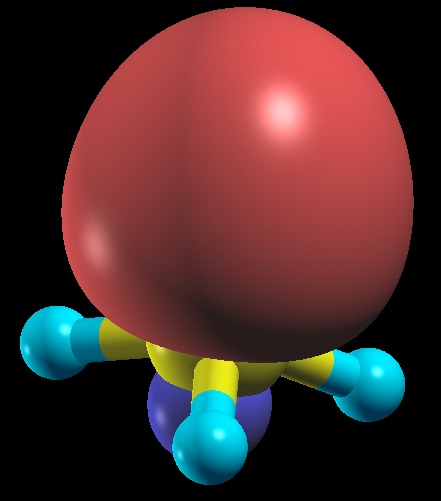}}
%\put(-51,116){\fbox{\includegraphics[height=1.8cm]{MLWFMethane.png}}}
\end{picture}
\caption{(Color on line) Difference, $E_\mathrm{SO}-E_\mathrm{NSO}$, between the band structure of chain of methane 
molecules calculated with, $E_\mathrm{SO}$, and without, $E_\mathrm{SO}$, considering SO interaction. The bands are 
labelled in increasing energy order without taking into account spin degeneracy. The left-hand side panels show results for 
the SO-DFT calculations performed with {\sc Siesta}, while the right-hand side one, those obtained from the MLWFs. 
The inset shows an isovalue plot of one of the four MLWFs with the red and blue surfaces denoting positive and negative 
isovalues, respectively. All the MLWFs have similar structure and they resemble those of the isolated methane molecule 
because of the small intermolecular chemical bonding owing to the large separation. } \label{fig:Chain_methane}
\end{figure}

As a first basic prototype of 1D organic molecular crystal we perform calculations for a periodic chain of 
methane molecules. We use a double-zeta polarized basis set and a LDA-relaxed unit cell length of 3.45~\AA\ 
(the cell contains only one molecule). Similarly to the previous case, the MLWFs are constructed over 
only the lowest 4 bands (8 when considering the spin degeneracy). When compared to the bands of the 
carbon chain, those of methane are much narrower. This is expected, since the bonding between the different 
molecules is small. In Fig.~\ref{fig:Chain_methane} we plot the difference between the eigenvalues (1D band 
structure) calculated with, $E_\mathrm{SO}$, and without, $E_\mathrm{NSO}$, including SO interaction. 

When SO interaction is included the spin-degeneracy is broken and one has now eight bands. These are labeled
as $\psi_m$ in Fig.~\ref{fig:Chain_methane} in increasing energy order. Again we find no SO split for the lowermost 
band and then a split, which is significantly smaller than that found in the case of the C chain. This is likely to originate 
from the crystal field of the C atoms in CH$_4$, which is different from that in the C chain (the C-C distance is different 
and there are additional C-H bonds). Again, as in the previous case, we find that our MLWFs procedure perfectly reproduces 
the SO-DFT band structure, indicating that in this case of weak SO interaction band truncation does not introduce 
any significant error.

\subsection{Triarylamine Chain}

\begin{figure}
\centering
\includegraphics[width=0.35\textwidth]{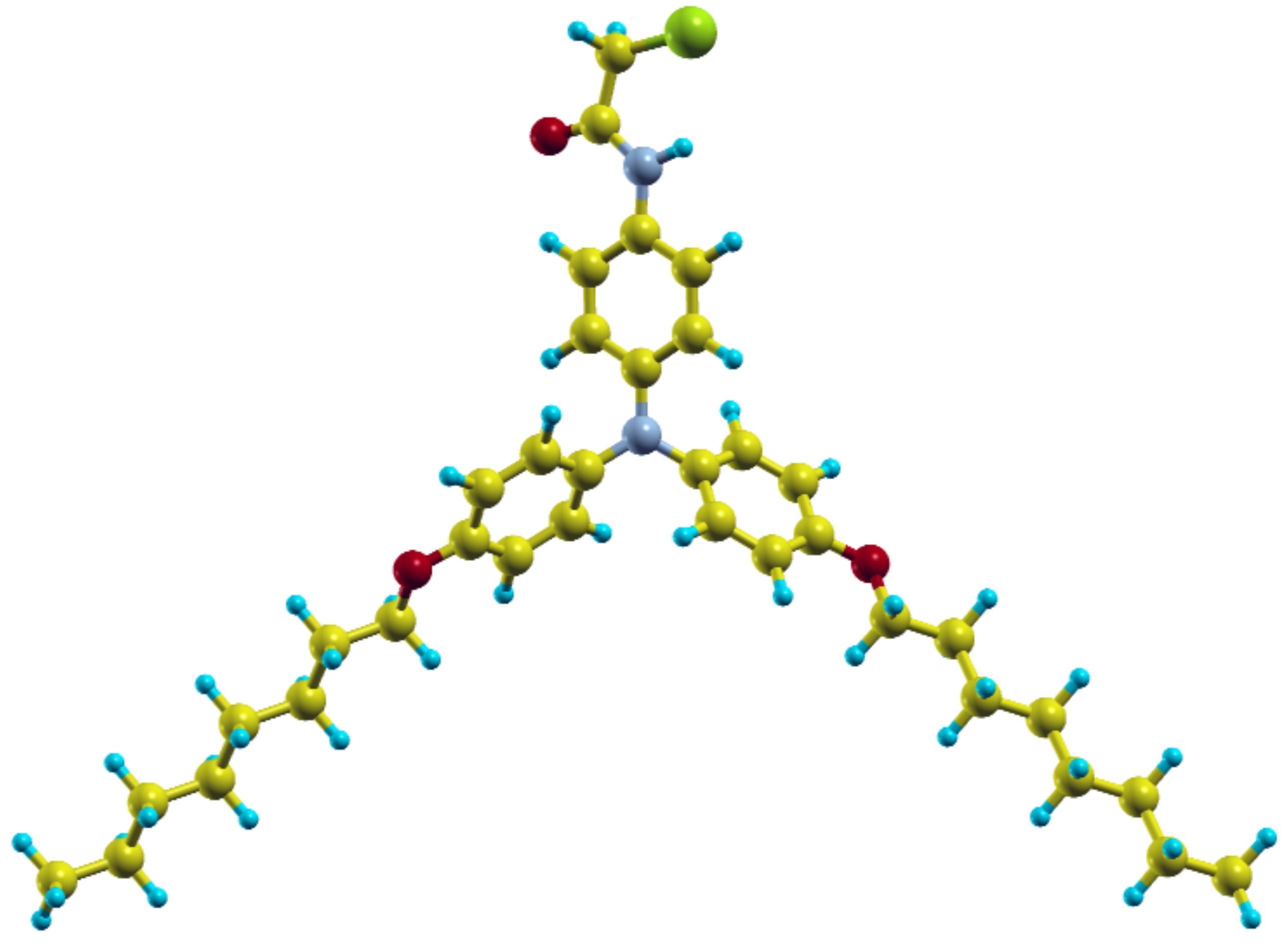}
\includegraphics[width=0.35\textwidth]{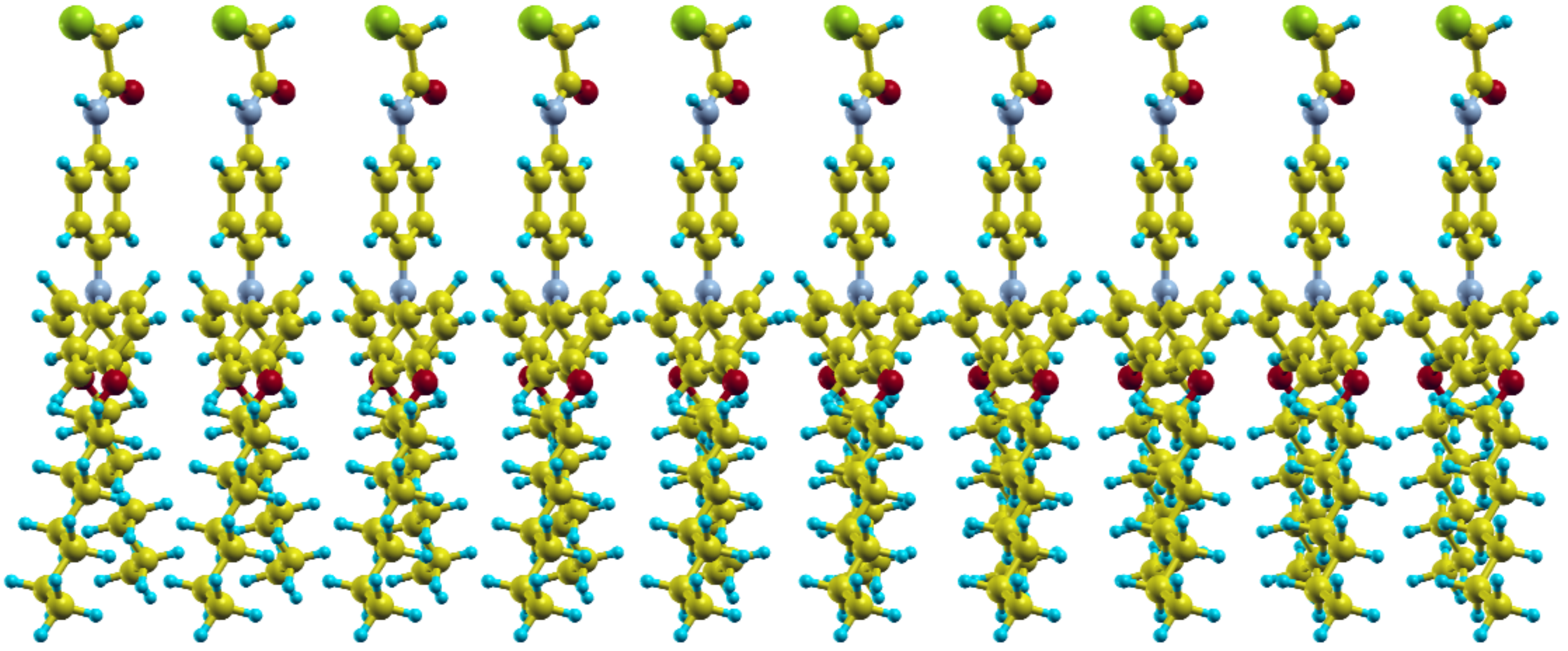}
\caption{(Colour on line) Structure of the triarylamine molecule (upper picture) and of the triarylamine-based nanowire
investigated here. The radicals associated to the triarylamine derivative are C$_8$H$_{17}$, H and Cl, respectively. 
Colour code: C=yellow, H=light blue, O=red, N=grey, Cl=green.}\label{fig:triarylamine_structure}
\end{figure}

Finally we perform calculations for a real system, namely for triarylamine-based molecular nanowires. These can be 
experimentally grown through a photo-self-assembly process from the liquid phase~\cite{ANIE:ANIE201001833}, 
and have been subject of numerous experimental and theoretical studies~\cite{B908802D,Vina}. In general, 
triarylamines can be used as materials for organic light emitting diodes, while their nanowire form appears to 
possess good transport and spin properties, making it a good platform for organic spintronics~\cite{Akin}. 
Triarylamine-based molecular nanowires self-assemble only when particular radicals are attached to the main
triarylamine backbone and here we consider the case of C$_8$H$_{17}$, H and Cl radicals, corresponding
to the precursor {\bf 1} of Ref.~[\onlinecite{ANIE:ANIE201001833}] (see upper panel in Fig.~\ref{fig:triarylamine_structure}). 
The nanowire then arranges in such a way to have the central N atoms aligned along the wire axis (see 
Fig.~\ref{fig:triarylamine_structure}). 

\begin{figure}
\centering
\includegraphics[width=0.30\textwidth]{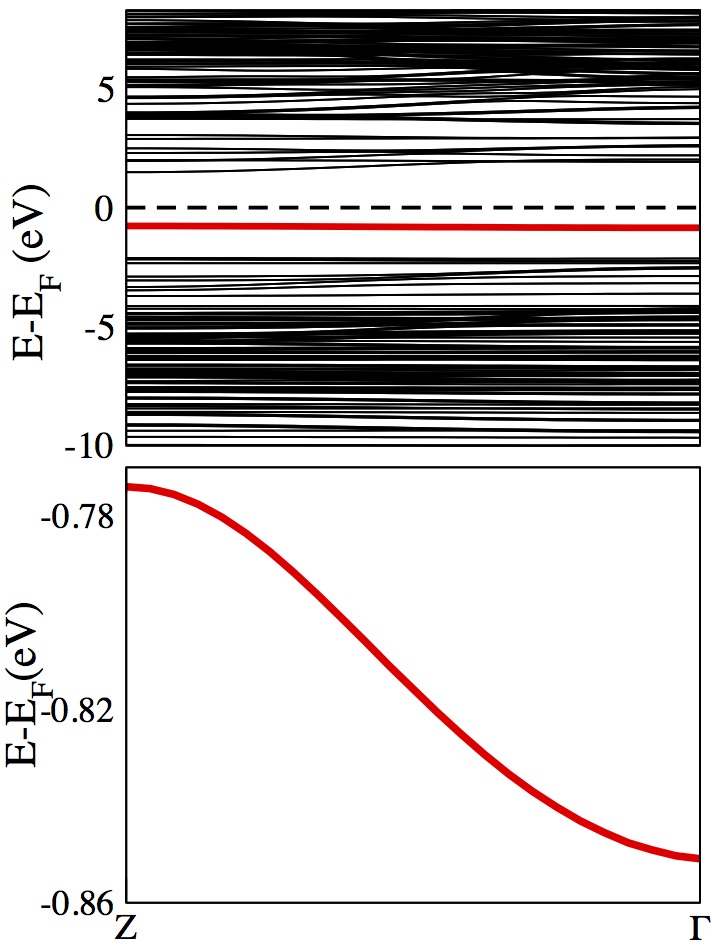}%
\caption{(Color on line) Band structure of the 1D triarylamine-based nanowire constructed with the precursor {\bf 1} 
of Ref.~[\onlinecite{ANIE:ANIE201001833}]. This is plotted over the 1D Brillouin zone (Z=$\pi/a$ with $a$ the
lattice parameter). The Fermi level is marked with a dashed black line and it is placed just above the HOMO-derived
valence band (in red). The lower panel is a magnification of the valence band. Note the bandwidth of about
100~meV and the fact that the band has a cosine shape, fingerprint of a single-orbital nearest-neighbour
tight-binding-like interaction. Only the HOMO band is considered when constructing the MLWFs.}\label{fig:BS_triarylamine}
\label{fig:triarylamine_BS}%
\end{figure}

In general self-assembled triarylamine-based molecular nanowires appear slightly $p$-doped so that charge
transport takes place in the HOMO-derived band. This is well isolated from the rest of the valence manifold 
and has a bandwidth of about 100~meV (see figure Fig.~\ref{fig:BS_triarylamine} for the band structure).
Such band is almost entirely localized on the $p_z$ orbital of the central N atoms ($p_z$ is along the wire
axis), a feature that has allowed us to construct a $p_z$-$sp^2$ model with the spin-orbit strength extracted from 
that of an equivalent mono-atomic N chain. The model was then used to calculate the temperature-dependent 
spin-diffusion length of such nanowires~\cite{C4CC01710B}. Here we wish to use our MLWFs method to extract 
the SO matrix elements of triarylamine-based molecular nanowires in their own chemical environment, i.e. 
without approximating the backbone with a N atomic chain.

For this system we use a 1D lattice with LDA-optimized lattice spacing of 4.8~\AA\ and run the DFT 
calculations with double-zeta polarized basis and the LDA functional. The MLWFs are constructed by using 
only the HOMO-derived valence band, i.e. we have a single spin-degenerate  Wannier orbital. We can then 
drop the band index and write the SO matrix elements as
\begin{multline}\label{equ82}
\begin{split}
& \bra{w_{\mathbf{0}}^{s_1}}\mathbf{V}_\mathrm{SO}\ket{w_{\mathbf{R}}^{s_2}}\\ 
&=\frac{V}{(2\pi)^3}\int d\mathbf{k} U^*(\mathbf{k})U(\mathbf{k})e^{-i\mathbf{k}.\mathbf{R}}\bra{\psi_{\mathbf{k}}^{s_1}}
\mathbf{V}_{SO}\ket{\psi_{\mathbf{k}}^{s_2}} \\
& =\frac{V}{(2\pi)^3}\int d\mathbf{k} e^{-i\mathbf{k}.\mathbf{R}}\bra{\psi_{\mathbf{k}}^{s_1}}
\mathbf{V}_\mathrm{SO}\ket{\psi_{\mathbf{k}}^{s_2}}\:,
\end{split}
\end{multline}
or in a discrete representation of the reciprocal space 
\begin{multline}\label{equ82}
\begin{split}
\bra{w_{\mathbf{0}}^{s_1}}\mathbf{V}_\mathrm{SO}\ket{w_{\mathbf{R}}^{s_2}} 
&=\frac{1}{N}\sum_{\mathbf{k}}e^{-i\mathbf{k}.\mathbf{R}}\bra{\psi_{\mathbf{k}}^{s_1}}
\mathbf{V}_\mathrm{SO}\ket{\psi_{\mathbf{k}}^{s_2}} \\
\end{split}
\end{multline}
where the second equality comes from the unitarity of the gauge transformation, $U(\mathbf{k})$. 

\begin{figure}
\centering
\includegraphics[width=0.48\textwidth]{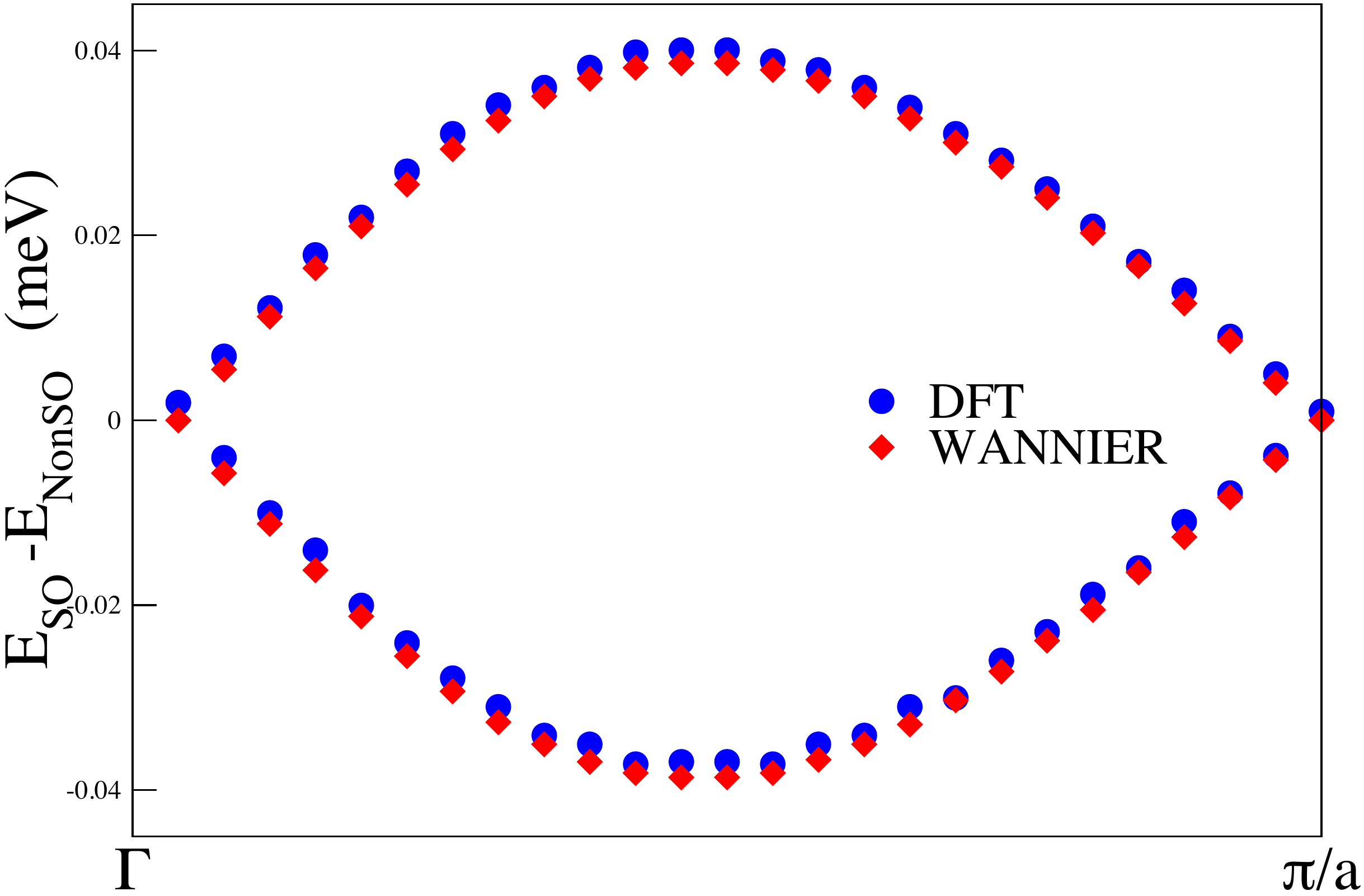}%
\caption{(Color on line) Plot of ($\rm{E_{SO}}-\rm{E_{NSO}}$) as a 
function of \textbf{k} in arbitrary unit over a brillouin zone for the highest occupied band of a 1-d chain of 
triarylamine derivatives. The blue and the red points correspond to calculations with {\sc Siesta} and 
{\sc Wannier90} respectively.}\label{fig:triarylamine_Bandstructure}
\label{fig:BSdiff_triarylamine}%
\end{figure}

In Fig.~\ref{fig:triarylamine_Bandstructure} we plot the difference between the band structure computed
by including SO interaction and those calculated without. Notably our MLWFs band structure is almost
identical to that computed directly with SO-DFT, again demonstrating both the accuracy of our method 
and the appropriateness of the drastic band truncation used here. 
In this particular case the SO band split is maximized half-way between the $\Gamma$ point and the edge of
the 1D Brillouin zone, where it takes a value of approximately 80~$\mu$eV. Clearly such split is orders of magnitude
smaller than the value that one can possibly calculate by a direct construction of the MLWFs from the SO-splitted
band structure. Note also that the SO split of the valence band is calculated here approximately a factor
ten smaller than that estimated previously for a N atomic chain~\cite{C4CC01710B}, indicating the importance
of the details of the chemical environment in these calculations.

\begin{figure}
        \centering
\subfloat{\label{fig:CoupMatElms_triarylamine}%
  \includegraphics[width=0.5\textwidth]{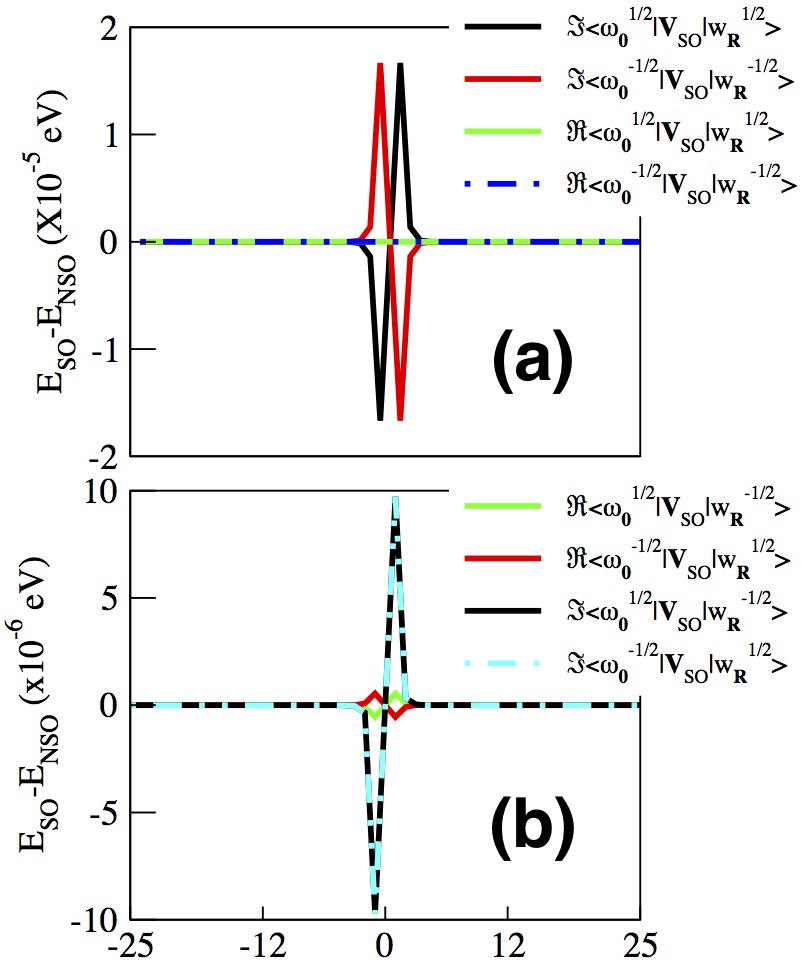}%
}\    
\caption{(Color on line) SO matrix elements of a triarylamine-based nanowire calculated with respect to the 
Wannier functions obtained from the HOMO band. Panels (a) and (b) correspond to matrix elements calculated 
between for same and different spins, respectively.}\label{fig:triarylamine_spinorbit}
\end{figure}

Finally we take a closer look at the calculated SO matrix elements. As mentioned earlier, in the {\sc Siesta} 
on-site approximation~\cite{0953-8984-18-34-012} only the matrix elements calculated over orbitals centred 
on the same atom do not vanish. As a consequence the components 
$\bra{w_{\mathbf{R}}^{s_1}}\mathbf{V}_\mathrm{SO}\ket{w_{\mathbf{R'}}^{s_2}}$ drop to zero as 
$|\mathbf{R}-\mathbf{R'}|$ gets large. This can be clearly appreciated in Fig.~\ref{fig:triarylamine_spinorbit}(a) 
and Fig.~\ref{fig:triarylamine_spinorbit}(b), where we plot the SO matrix elements for same and different spins, 
respectively. 

From Fig.~\ref{fig:triarylamine_spinorbit}(a) we can observe that 
$\Re \bra{w_{\mathbf{0}}^{s_1}}\mathbf{V}_\mathrm{SO}\ket{w_{\mathbf{R}}^{s_1}}$ vanishes for all \textbf{R}. 
This can be understood in the following way. In general any expectation value of $\mathbf{V}_{\rm{SO}}$,
$\bra{\psi_{\mathbf{k}}^{s}}\mathbf{V}_\mathrm{SO}\ket{\psi^{s}_{\mathbf{k}}}$, has to be real. This is in 
fact anti-symmetric with respect to \textbf{k}, i.e we have 
$\bra{\psi_{\mathbf{0}+\mathbf{k}}^{s}}\mathbf{V}_{\rm{SO}}\ket{\psi^{s}_{\mathbf{0}+\mathbf{k}}}=-\bra{\psi_{\mathbf{0}-\mathbf{k}}^{s}}\mathbf{V}_{\rm{SO}}\ket{\psi^{s}_{\mathbf{0}-\mathbf{k}}}$, where 
$\mathbf{k}=\mathbf{0}$ denotes the $\Gamma$ point of the Brillouin zone. 
Additionally, $e^{i\mathbf{k}\cdot\mathbf{R}}$ satisfies the relation 
$e^{i(\mathbf{0}+\mathbf{k})\cdot\mathbf{R}}=\left[e^{i(\mathbf{0}-\mathbf{k})\cdot\mathbf{R}}\right]^*$. 
Hence, by performing the \textbf{k}-sum over first Brillouin zone we can write
\begin{multline}\label{equ82}
\begin{split}
\Re \bra{w_{\mathbf{0}}^{s_1}}\mathbf{V}_{\rm{SO}}\ket{w_{\mathbf{R}}^{s_1}}=\Re \sum_{\mathbf{k}} e^{-i\mathbf{k}.\mathbf{R}}\bra{\psi_{\mathbf{k}}^{s_1}}\mathbf{V}_{\rm{SO}}\ket{\psi_{\mathbf{k}}^{s_1}}=0\:,
\end{split}
\end{multline}
where $\bra{w_{\mathbf{0}}^{s_1}}\mathbf{V}_{\rm{SO}}\ket{w_{\mathbf{0}}^{s_1}}$ is the expectation value of 
$\mathbf{V}_{\rm{SO}}$ and must be real. This implies
\begin{multline}\label{equ82}
\begin{split}
 \bra{w_{\mathbf{0}}^{s_1}}\mathbf{V}_{\rm{SO}}\ket{w_{\mathbf{0}}^{s_1}}=0\:.
\end{split}
\end{multline}

We can also see from Fig.~\ref{fig:triarylamine_spinorbit}(b) that for triarylamine the matrix elements 
$\bra{w_{\mathbf{R}}^{s_1}}\mathbf{V}_\mathrm{SO}\ket{w_{\mathbf{R}}^{s_2}}$ are almost zero for $s_1 \neq s_2$. 
This follows directly from Eq.~(\ref{equ20}). In fact in the particular case of triarylamine nanowires
the Wannier functions are constructed from one band only. As such, in order to have a non-zero matrix element,
$\bra{w_{\mathbf{R}}^{s_1}}\mathbf{V}_\mathrm{SO}\ket{w_{\mathbf{R}}^{s_2}}$, we must have non-zero values 
for $\bra{w_{\mathbf{R}}}\hat{L}_{\pm}\ket{w_{\mathbf{R}}}$. Therefore, the band under consideration must contain 
an appreciable mix of components of both the $\ket{l,p}$ and $\ket{l,p+1}$ complex spherical harmonics for some $l$ 
and $p$. As mentioned earlier, the triarylamine HOMO band is composed mostly of $p_z$ N orbitals. Hence, it has 
to be expected that the $\bra{w_{\mathbf{R}}^{s_1}}\mathbf{V}_\mathrm{SO}\ket{w_{\mathbf{R}}^{s_2}}$ matrix
elements are small.

\section{conclusion}

We have presented an accurate method for obtaining the SO matrix elements between the MLWFs constructed 
in absence of SO coupling. Our procedure, implemented within the atomic-orbital-based DFT code {\sc Siesta},
allows one to avoid the construction of the Wannier functions over the SO-split band structure. In some cases,
in particular for organic crystals, such splits are tiny and a direct construction is numerically impossible. The method
is then put to the test for a number of materials systems, going from isolated molecules, to atomic nanowires, to
1D molecular crystals. When the entire band manifold is used for constructing the MLWFs the mapping between
Bloch and Wannier orbitals is exact and the method can be used for both light and heavy elements. In contrast
for weak spin-orbit interaction one can construct the MLWFs on a subset of the states in the band structures
without any loss of accuracy. As such our scheme appears as an important tool for constructing effective spin
Hamiltonians for organic materials to be used as input in a multiscale approach to the their thermodynamical 
properties.

\section*{ackowledgement}

This work is supported by the European Research Council, Quest project. Computational resources have 
been provided by the supercomputer facilities at the Trinity Center for High Performance Computing (TCHPC) 
and at the Irish Center for High End Computing (ICHEC). Additionally, the authors would like to thank Ivan Rungger 
and Carlo Motta for helpful discussions and Akinlolu Akande for providing the structure of the triarylamine-based 
nanowire.


\begin{thebibliography}{42}%
\makeatletter
\providecommand \@ifxundefined [1]{%
 \@ifx{#1\undefined}
}%
\providecommand \@ifnum [1]{%
 \ifnum #1\expandafter \@firstoftwo
 \else \expandafter \@secondoftwo
 \fi
}%
\providecommand \@ifx [1]{%
 \ifx #1\expandafter \@firstoftwo
 \else \expandafter \@secondoftwo
 \fi
}%
\providecommand \natexlab [1]{#1}%
\providecommand \enquote  [1]{``#1''}%
\providecommand \bibnamefont  [1]{#1}%
\providecommand \bibfnamefont [1]{#1}%
\providecommand \citenamefont [1]{#1}%
\providecommand \href@noop [0]{\@secondoftwo}%
\providecommand \href [0]{\begingroup \@sanitize@url \@href}%
\providecommand \@href[1]{\@@startlink{#1}\@@href}%
\providecommand \@@href[1]{\endgroup#1\@@endlink}%
\providecommand \@sanitize@url [0]{\catcode `\\12\catcode `\$12\catcode
  `\&12\catcode `\#12\catcode `\^12\catcode `\_12\catcode `\%12\relax}%
\providecommand \@@startlink[1]{}%
\providecommand \@@endlink[0]{}%
\providecommand \url  [0]{\begingroup\@sanitize@url \@url }%
\providecommand \@url [1]{\endgroup\@href {#1}{\urlprefix }}%
\providecommand \urlprefix  [0]{URL }%
\providecommand \Eprint [0]{\href }%
\providecommand \doibase [0]{http://dx.doi.org/}%
\providecommand \selectlanguage [0]{\@gobble}%
\providecommand \bibinfo  [0]{\@secondoftwo}%
\providecommand \bibfield  [0]{\@secondoftwo}%
\providecommand \translation [1]{[#1]}%
\providecommand \BibitemOpen [0]{}%
\providecommand \bibitemStop [0]{}%
\providecommand \bibitemNoStop [0]{.\EOS\space}%
\providecommand \EOS [0]{\spacefactor3000\relax}%
\providecommand \BibitemShut  [1]{\csname bibitem#1\endcsname}%
\let\auto@bib@innerbib\@empty
%</preamble>
\bibitem [{\citenamefont {Wolf}\ \emph {et~al.}(2001)\citenamefont {Wolf},
  \citenamefont {Awschalom}, \citenamefont {Buhrman}, \citenamefont {Daughton},
  \citenamefont {von Moln{\'a}r}, \citenamefont {Roukes}, \citenamefont
  {Chtchelkanova},\ and\ \citenamefont {Treger}}]{Wolf1488}%
  \BibitemOpen
  \bibfield  {author} {\bibinfo {author} {\bibfnamefont {S.~A.}\ \bibnamefont
  {Wolf}}, \bibinfo {author} {\bibfnamefont {D.~D.}\ \bibnamefont {Awschalom}},
  \bibinfo {author} {\bibfnamefont {R.~A.}\ \bibnamefont {Buhrman}}, \bibinfo
  {author} {\bibfnamefont {J.~M.}\ \bibnamefont {Daughton}}, \bibinfo {author}
  {\bibfnamefont {S.}~\bibnamefont {von Moln{\'a}r}}, \bibinfo {author}
  {\bibfnamefont {M.~L.}\ \bibnamefont {Roukes}}, \bibinfo {author}
  {\bibfnamefont {A.~Y.}\ \bibnamefont {Chtchelkanova}}, \ and\ \bibinfo
  {author} {\bibfnamefont {D.~M.}\ \bibnamefont {Treger}},\ }\href {\doibase
  10.1126/science.1065389} {\bibfield  {journal} {\bibinfo  {journal}
  {Science}\ }\textbf {\bibinfo {volume} {294}},\ \bibinfo {pages} {1488}
  (\bibinfo {year} {2001})}\BibitemShut {NoStop}%
\bibitem [{\citenamefont {Ornes}(2013)}]{Ornes05032013}%
  \BibitemOpen
  \bibfield  {author} {\bibinfo {author} {\bibfnamefont {S.}~\bibnamefont
  {Ornes}},\ }\href {\doibase 10.1073/pnas.1302494110} {\bibfield  {journal}
  {\bibinfo  {journal} {Proc. Natl. Acad. Sci. USA}\ }\textbf {\bibinfo
  {volume} {110}},\ \bibinfo {pages} {3710} (\bibinfo {year}
  {2013})}\BibitemShut {NoStop}%
\bibitem [{\citenamefont {Behin-Aein}\ \emph {et~al.}(2010)\citenamefont
  {Behin-Aein}, \citenamefont {Datta}, \citenamefont {Salahuddin},\ and\
  \citenamefont {Datta}}]{DattaSLogic}%
  \BibitemOpen
  \bibfield  {author} {\bibinfo {author} {\bibfnamefont {B.}~\bibnamefont
  {Behin-Aein}}, \bibinfo {author} {\bibfnamefont {D.}~\bibnamefont {Datta}},
  \bibinfo {author} {\bibfnamefont {S.}~\bibnamefont {Salahuddin}}, \ and\
  \bibinfo {author} {\bibfnamefont {S.}~\bibnamefont {Datta}},\ }\href@noop {}
  {\bibfield  {journal} {\bibinfo  {journal} {Nature Nanotech.}\ }\textbf
  {\bibinfo {volume} {5}},\ \bibinfo {pages} {266} (\bibinfo {year}
  {2010})}\BibitemShut {NoStop}%
\bibitem [{\citenamefont {{Prinz}}\ and\ \citenamefont
  {{Hathaway}}(1995)}]{1995PhT}%
  \BibitemOpen
  \bibfield  {author} {\bibinfo {author} {\bibfnamefont {G.}~\bibnamefont
  {{Prinz}}}\ and\ \bibinfo {author} {\bibfnamefont {K.}~\bibnamefont
  {{Hathaway}}},\ }\href {\doibase 10.1063/1.881446} {\bibfield  {journal}
  {\bibinfo  {journal} {Phys Today}\ }\textbf {\bibinfo {volume} {48}},\
  \bibinfo {pages} {24} (\bibinfo {year} {1995})}\BibitemShut {NoStop}%
\bibitem [{\citenamefont {Awschalom}\ and\ \citenamefont
  {Flatt{\'e}}(2007)}]{Awschalom}%
  \BibitemOpen
  \bibfield  {author} {\bibinfo {author} {\bibfnamefont {D.~D.}\ \bibnamefont
  {Awschalom}}\ and\ \bibinfo {author} {\bibfnamefont {M.~E.}\ \bibnamefont
  {Flatt{\'e}}},\ }\href {\doibase doi:10.1166/jctn.2006.003} {\bibfield
  {journal} {\bibinfo  {journal} {Nature Physics}\ }\textbf {\bibinfo {volume}
  {3}},\ \bibinfo {pages} {153} (\bibinfo {year} {2007})}\BibitemShut {NoStop}%
\bibitem [{\citenamefont {\ifmmode \check{Z}\else
  \v{Z}\fi{}uti\ifmmode~\acute{c}\else \'{c}\fi{}}\ and\ \citenamefont
  {Fuhrer}(2005)}]{Igor}%
  \BibitemOpen
  \bibfield  {author} {\bibinfo {author} {\bibfnamefont {I.}~\bibnamefont
  {\ifmmode \check{Z}\else \v{Z}\fi{}uti\ifmmode~\acute{c}\else \'{c}\fi{}}}\
  and\ \bibinfo {author} {\bibfnamefont {M.}~\bibnamefont {Fuhrer}},\
  }\href@noop {} {\bibfield  {journal} {\bibinfo  {journal} {Nature Physics}\
  }\textbf {\bibinfo {volume} {1}},\ \bibinfo {pages} {85} (\bibinfo {year}
  {2005})}\BibitemShut {NoStop}%
\bibitem [{\citenamefont {Horowitz}(2006)}]{HoroGill}%
  \BibitemOpen
  \bibfield  {author} {\bibinfo {author} {\bibfnamefont {G.}~\bibnamefont
  {Horowitz}},\ }\href@noop {} {\emph {\bibinfo {title} {Organic
  Transistors}}},\ edited by\ \bibinfo {editor} {\bibfnamefont
  {H.}~\bibnamefont {Klauk}}\ (\bibinfo  {publisher} {Wiley-VCH Verlag GmbH \&
  Co. KGaA},\ \bibinfo {year} {2006})\BibitemShut {NoStop}%
\bibitem [{\citenamefont {Joachim}\ \emph {et~al.}(2000)\citenamefont
  {Joachim}, \citenamefont {Gimzewski},\ and\ \citenamefont {Aviram}}]{JOUR1}%
  \BibitemOpen
  \bibfield  {author} {\bibinfo {author} {\bibfnamefont {C.}~\bibnamefont
  {Joachim}}, \bibinfo {author} {\bibfnamefont {J.~K.}\ \bibnamefont
  {Gimzewski}}, \ and\ \bibinfo {author} {\bibfnamefont {A.}~\bibnamefont
  {Aviram}},\ }\href@noop {} {\bibfield  {journal} {\bibinfo  {journal}
  {Nature}\ }\textbf {\bibinfo {volume} {408}},\ \bibinfo {pages} {541}
  (\bibinfo {year} {2000})}\BibitemShut {NoStop}%
\bibitem [{\citenamefont {Dediu}\ \emph {et~al.}(2002)\citenamefont {Dediu},
  \citenamefont {Murgia}, \citenamefont {Matacotta}, \citenamefont {Taliani},\
  and\ \citenamefont {Barbanera}}]{Dediu2002181}%
  \BibitemOpen
  \bibfield  {author} {\bibinfo {author} {\bibfnamefont {V.}~\bibnamefont
  {Dediu}}, \bibinfo {author} {\bibfnamefont {M.}~\bibnamefont {Murgia}},
  \bibinfo {author} {\bibfnamefont {F.}~\bibnamefont {Matacotta}}, \bibinfo
  {author} {\bibfnamefont {C.}~\bibnamefont {Taliani}}, \ and\ \bibinfo
  {author} {\bibfnamefont {S.}~\bibnamefont {Barbanera}},\ }\href {\doibase
  http://dx.doi.org/10.1016/S0038-1098(02)00090-X} {\bibfield  {journal}
  {\bibinfo  {journal} {Solid State Commun.}\ }\textbf {\bibinfo {volume}
  {122}},\ \bibinfo {pages} {181 } (\bibinfo {year} {2002})}\BibitemShut
  {NoStop}%
\bibitem [{\citenamefont {Xiong}\ \emph {et~al.}(2004)\citenamefont {Xiong},
  \citenamefont {Wu}, \citenamefont {Valy~Vardeny},\ and\ \citenamefont
  {Shi}}]{JOUR}%
  \BibitemOpen
  \bibfield  {author} {\bibinfo {author} {\bibfnamefont {Z.~H.}\ \bibnamefont
  {Xiong}}, \bibinfo {author} {\bibfnamefont {D.}~\bibnamefont {Wu}}, \bibinfo
  {author} {\bibfnamefont {Z.}~\bibnamefont {Valy~Vardeny}}, \ and\ \bibinfo
  {author} {\bibfnamefont {J.}~\bibnamefont {Shi}},\ }\href@noop {} {\bibfield
  {journal} {\bibinfo  {journal} {Nature}\ }\textbf {\bibinfo {volume} {427}},\
  \bibinfo {pages} {821} (\bibinfo {year} {2004})}\BibitemShut {NoStop}%
\bibitem [{\citenamefont {Sanvito}(2011)}]{C1CS15047B}%
  \BibitemOpen
  \bibfield  {author} {\bibinfo {author} {\bibfnamefont {S.}~\bibnamefont
  {Sanvito}},\ }\href {\doibase 10.1039/C1CS15047B} {\bibfield  {journal}
  {\bibinfo  {journal} {Chem. Soc. Rev.}\ }\textbf {\bibinfo {volume} {40}},\
  \bibinfo {pages} {3336} (\bibinfo {year} {2011})}\BibitemShut {NoStop}%
\bibitem [{\citenamefont {Chiang}\ \emph {et~al.}(1977)\citenamefont {Chiang},
  \citenamefont {Fincher}, \citenamefont {Park}, \citenamefont {Heeger},
  \citenamefont {Shirakawa}, \citenamefont {Louis}, \citenamefont {Gau},\ and\
  \citenamefont {MacDiarmid}}]{PhysRevLett.39.1098}%
  \BibitemOpen
  \bibfield  {author} {\bibinfo {author} {\bibfnamefont {C.~K.}\ \bibnamefont
  {Chiang}}, \bibinfo {author} {\bibfnamefont {C.~R.}\ \bibnamefont {Fincher}},
  \bibinfo {author} {\bibfnamefont {Y.~W.}\ \bibnamefont {Park}}, \bibinfo
  {author} {\bibfnamefont {A.~J.}\ \bibnamefont {Heeger}}, \bibinfo {author}
  {\bibfnamefont {H.}~\bibnamefont {Shirakawa}}, \bibinfo {author}
  {\bibfnamefont {E.~J.}\ \bibnamefont {Louis}}, \bibinfo {author}
  {\bibfnamefont {S.~C.}\ \bibnamefont {Gau}}, \ and\ \bibinfo {author}
  {\bibfnamefont {A.~G.}\ \bibnamefont {MacDiarmid}},\ }\href {\doibase
  10.1103/PhysRevLett.39.1098} {\bibfield  {journal} {\bibinfo  {journal}
  {Phys. Rev. Lett.}\ }\textbf {\bibinfo {volume} {39}},\ \bibinfo {pages}
  {1098} (\bibinfo {year} {1977})}\BibitemShut {NoStop}%
\bibitem [{\citenamefont {Forrest}(2004)}]{Forrest}%
  \BibitemOpen
  \bibfield  {author} {\bibinfo {author} {\bibfnamefont {S.~R.}\ \bibnamefont
  {Forrest}},\ }\href@noop {} {\bibfield  {journal} {\bibinfo  {journal}
  {Nature}\ }\textbf {\bibinfo {volume} {428}},\ \bibinfo {pages} {911}
  (\bibinfo {year} {2004})}\BibitemShut {NoStop}%
\bibitem [{\citenamefont {Sanvito}\ and\ \citenamefont
  {Rocha}(2006)}]{Sanvito}%
  \BibitemOpen
  \bibfield  {author} {\bibinfo {author} {\bibfnamefont {S.}~\bibnamefont
  {Sanvito}}\ and\ \bibinfo {author} {\bibfnamefont {A.~R.}\ \bibnamefont
  {Rocha}},\ }\href@noop {} {\bibfield  {journal} {\bibinfo  {journal} {J.
  Comput. Theor. Nanosci.}\ }\textbf {\bibinfo {volume} {3}},\ \bibinfo {pages}
  {624} (\bibinfo {year} {2006})}\BibitemShut {NoStop}%
\bibitem [{\citenamefont {Pramanik}\ \emph {et~al.}(2007)\citenamefont
  {Pramanik}, \citenamefont {Stefanita}, \citenamefont {Patibandla},
  \citenamefont {Bandyopadhyay}, \citenamefont {Garre}, \citenamefont {N.},\
  and\ \citenamefont {Cahay}}]{Pramanik}%
  \BibitemOpen
  \bibfield  {author} {\bibinfo {author} {\bibfnamefont {S.}~\bibnamefont
  {Pramanik}}, \bibinfo {author} {\bibfnamefont {C.~G.}\ \bibnamefont
  {Stefanita}}, \bibinfo {author} {\bibfnamefont {S.}~\bibnamefont
  {Patibandla}}, \bibinfo {author} {\bibfnamefont {S.}~\bibnamefont
  {Bandyopadhyay}}, \bibinfo {author} {\bibfnamefont {K.}~\bibnamefont
  {Garre}}, \bibinfo {author} {\bibfnamefont {H.}~\bibnamefont {N.}}, \ and\
  \bibinfo {author} {\bibfnamefont {M.}~\bibnamefont {Cahay}},\ }\href@noop {}
  {\bibfield  {journal} {\bibinfo  {journal} {Nature Nanotech.}\ }\textbf
  {\bibinfo {volume} {2}},\ \bibinfo {pages} {216} (\bibinfo {year}
  {2007})}\BibitemShut {NoStop}%
\bibitem [{\citenamefont {Tsukagoshi}\ \emph {et~al.}(1999)\citenamefont
  {Tsukagoshi}, \citenamefont {Alphenaar},\ and\ \citenamefont
  {Ago}}]{Tsukagoshi}%
  \BibitemOpen
  \bibfield  {author} {\bibinfo {author} {\bibfnamefont {K.}~\bibnamefont
  {Tsukagoshi}}, \bibinfo {author} {\bibfnamefont {B.~W.}\ \bibnamefont
  {Alphenaar}}, \ and\ \bibinfo {author} {\bibfnamefont {H.}~\bibnamefont
  {Ago}},\ }\href@noop {} {\bibfield  {journal} {\bibinfo  {journal} {Nature}\
  }\textbf {\bibinfo {volume} {401}},\ \bibinfo {pages} {572} (\bibinfo {year}
  {1999})}\BibitemShut {NoStop}%
\bibitem [{\citenamefont {Szulczewski}\ \emph {et~al.}(2009)\citenamefont
  {Szulczewski}, \citenamefont {Sanvito},\ and\ \citenamefont
  {Coey}}]{SCS2009}%
  \BibitemOpen
  \bibfield  {author} {\bibinfo {author} {\bibfnamefont {G.}~\bibnamefont
  {Szulczewski}}, \bibinfo {author} {\bibfnamefont {S.}~\bibnamefont
  {Sanvito}}, \ and\ \bibinfo {author} {\bibfnamefont {J.~M.~D.}\ \bibnamefont
  {Coey}},\ }\href@noop {} {\bibfield  {journal} {\bibinfo  {journal} {Nature
  Materials}\ }\textbf {\bibinfo {volume} {8}},\ \bibinfo {pages} {693}
  (\bibinfo {year} {2009})}\BibitemShut {NoStop}%
\bibitem [{\citenamefont {Dieny}\ \emph {et~al.}(1991)\citenamefont {Dieny},
  \citenamefont {Speriosu}, \citenamefont {Gurney}, \citenamefont {Parkin},
  \citenamefont {Wilhoit}, \citenamefont {Roche}, \citenamefont {Metin},
  \citenamefont {Peterson},\ and\ \citenamefont {Nadimi}}]{SpinValve}%
  \BibitemOpen
  \bibfield  {author} {\bibinfo {author} {\bibfnamefont {B.}~\bibnamefont
  {Dieny}}, \bibinfo {author} {\bibfnamefont {V.}~\bibnamefont {Speriosu}},
  \bibinfo {author} {\bibfnamefont {B.}~\bibnamefont {Gurney}}, \bibinfo
  {author} {\bibfnamefont {S.}~\bibnamefont {Parkin}}, \bibinfo {author}
  {\bibfnamefont {D.}~\bibnamefont {Wilhoit}}, \bibinfo {author} {\bibfnamefont
  {K.}~\bibnamefont {Roche}}, \bibinfo {author} {\bibfnamefont
  {S.}~\bibnamefont {Metin}}, \bibinfo {author} {\bibfnamefont
  {D.}~\bibnamefont {Peterson}}, \ and\ \bibinfo {author} {\bibfnamefont
  {S.}~\bibnamefont {Nadimi}},\ }\href {\doibase
  http://dx.doi.org/10.1016/0304-8853(91)90311-W} {\bibfield  {journal}
  {\bibinfo  {journal} {Journal of Magnetism and Magnetic Materials}\ }\textbf
  {\bibinfo {volume} {93}},\ \bibinfo {pages} {101 } (\bibinfo {year}
  {1991})}\BibitemShut {NoStop}%
\bibitem [{\citenamefont {\ifmmode \check{Z}\else
  \v{Z}\fi{}uti\ifmmode~\acute{c}\else \'{c}\fi{}}\ \emph
  {et~al.}(2004)\citenamefont {\ifmmode \check{Z}\else
  \v{Z}\fi{}uti\ifmmode~\acute{c}\else \'{c}\fi{}}, \citenamefont {Fabian},\
  and\ \citenamefont {Das~Sarma}}]{RevModPhys.76.323}%
  \BibitemOpen
  \bibfield  {author} {\bibinfo {author} {\bibfnamefont {I.}~\bibnamefont
  {\ifmmode \check{Z}\else \v{Z}\fi{}uti\ifmmode~\acute{c}\else \'{c}\fi{}}},
  \bibinfo {author} {\bibfnamefont {J.}~\bibnamefont {Fabian}}, \ and\ \bibinfo
  {author} {\bibfnamefont {S.}~\bibnamefont {Das~Sarma}},\ }\href {\doibase
  10.1103/RevModPhys.76.323} {\bibfield  {journal} {\bibinfo  {journal} {Rev.
  Mod. Phys.}\ }\textbf {\bibinfo {volume} {76}},\ \bibinfo {pages} {323}
  (\bibinfo {year} {2004})}\BibitemShut {NoStop}%
\bibitem [{\citenamefont {Krinichnyi}(2000)}]{Impurity}%
  \BibitemOpen
  \bibfield  {author} {\bibinfo {author} {\bibfnamefont {V.~I.}\ \bibnamefont
  {Krinichnyi}},\ }\href@noop {} {\bibfield  {journal} {\bibinfo  {journal}
  {Synth. Met.}\ }\textbf {\bibinfo {volume} {108}},\ \bibinfo {pages} {173}
  (\bibinfo {year} {2000})}\BibitemShut {NoStop}%
\bibitem [{\citenamefont {Bandyopadhyay}(2010)}]{PhysRevB.81.153202}%
  \BibitemOpen
  \bibfield  {author} {\bibinfo {author} {\bibfnamefont {S.}~\bibnamefont
  {Bandyopadhyay}},\ }\href {\doibase 10.1103/PhysRevB.81.153202} {\bibfield
  {journal} {\bibinfo  {journal} {Phys. Rev. B}\ }\textbf {\bibinfo {volume}
  {81}},\ \bibinfo {pages} {153202} (\bibinfo {year} {2010})}\BibitemShut
  {NoStop}%
\bibitem [{\citenamefont {Drew}\ \emph {et~al.}(2009)\citenamefont {Drew} \emph
  {et~al.}}]{Drew}%
  \BibitemOpen
  \bibfield  {author} {\bibinfo {author} {\bibfnamefont {A.~J.}\ \bibnamefont
  {Drew}} \emph {et~al.},\ }\href@noop {} {\bibfield  {journal} {\bibinfo
  {journal} {Nature Materials}\ }\textbf {\bibinfo {volume} {8}},\ \bibinfo
  {pages} {109} (\bibinfo {year} {2009})}\BibitemShut {NoStop}%
\bibitem [{\citenamefont {Wang}\ \emph {et~al.}(2007)\citenamefont {Wang},
  \citenamefont {Yang}, \citenamefont {Vardeny},\ and\ \citenamefont
  {Li}}]{PhysRevB.75.245324}%
  \BibitemOpen
  \bibfield  {author} {\bibinfo {author} {\bibfnamefont {F.~J.}\ \bibnamefont
  {Wang}}, \bibinfo {author} {\bibfnamefont {C.~G.}\ \bibnamefont {Yang}},
  \bibinfo {author} {\bibfnamefont {Z.~V.}\ \bibnamefont {Vardeny}}, \ and\
  \bibinfo {author} {\bibfnamefont {X.~G.}\ \bibnamefont {Li}},\ }\href
  {\doibase 10.1103/PhysRevB.75.245324} {\bibfield  {journal} {\bibinfo
  {journal} {Phys. Rev. B}\ }\textbf {\bibinfo {volume} {75}},\ \bibinfo
  {pages} {245324} (\bibinfo {year} {2007})}\BibitemShut {NoStop}%
\bibitem [{\citenamefont {Dediu}\ \emph {et~al.}(2008)\citenamefont {Dediu},
  \citenamefont {Hueso}, \citenamefont {Bergenti}, \citenamefont {Riminucci},
  \citenamefont {Borgatti}, \citenamefont {Graziosi}, \citenamefont {Newby},
  \citenamefont {Casoli}, \citenamefont {De~Jong}, \citenamefont {Taliani},\
  and\ \citenamefont {Zhan}}]{PhysRevB.78.115203}%
  \BibitemOpen
  \bibfield  {author} {\bibinfo {author} {\bibfnamefont {V.}~\bibnamefont
  {Dediu}}, \bibinfo {author} {\bibfnamefont {L.~E.}\ \bibnamefont {Hueso}},
  \bibinfo {author} {\bibfnamefont {I.}~\bibnamefont {Bergenti}}, \bibinfo
  {author} {\bibfnamefont {A.}~\bibnamefont {Riminucci}}, \bibinfo {author}
  {\bibfnamefont {F.}~\bibnamefont {Borgatti}}, \bibinfo {author}
  {\bibfnamefont {P.}~\bibnamefont {Graziosi}}, \bibinfo {author}
  {\bibfnamefont {C.}~\bibnamefont {Newby}}, \bibinfo {author} {\bibfnamefont
  {F.}~\bibnamefont {Casoli}}, \bibinfo {author} {\bibfnamefont {M.~P.}\
  \bibnamefont {De~Jong}}, \bibinfo {author} {\bibfnamefont {C.}~\bibnamefont
  {Taliani}}, \ and\ \bibinfo {author} {\bibfnamefont {Y.}~\bibnamefont
  {Zhan}},\ }\href {\doibase 10.1103/PhysRevB.78.115203} {\bibfield  {journal}
  {\bibinfo  {journal} {Phys. Rev. B}\ }\textbf {\bibinfo {volume} {78}},\
  \bibinfo {pages} {115203} (\bibinfo {year} {2008})}\BibitemShut {NoStop}%
\bibitem [{\citenamefont {Cohen-Tannoudji}\ \emph {et~al.}(1977)\citenamefont
  {Cohen-Tannoudji}, \citenamefont {Diu},\ and\ \citenamefont
  {Lalo{\.e}}}]{cohen1977quantum}%
  \BibitemOpen
  \bibfield  {author} {\bibinfo {author} {\bibfnamefont {C.}~\bibnamefont
  {Cohen-Tannoudji}}, \bibinfo {author} {\bibfnamefont {B.}~\bibnamefont
  {Diu}}, \ and\ \bibinfo {author} {\bibfnamefont {F.}~\bibnamefont
  {Lalo{\.e}}},\ }\href {https://books.google.ie/books?id=BT5RAAAAMAAJ} {\emph
  {\bibinfo {title} {Quantum mechanics. 2}}},\ Textbook physics\ (\bibinfo
  {publisher} {John Wiley \& Sons},\ \bibinfo {year} {1977})\BibitemShut
  {NoStop}%
\bibitem [{\citenamefont {Troisi}(2011)}]{Troisi}%
  \BibitemOpen
  \bibfield  {author} {\bibinfo {author} {\bibfnamefont {A.}~\bibnamefont
  {Troisi}},\ }\href@noop {} {\bibfield  {journal} {\bibinfo  {journal} {J.
  Chem. Phys.}\ }\textbf {\bibinfo {volume} {134}},\ \bibinfo {eid} {034702}
  (\bibinfo {year} {2011})}\BibitemShut {NoStop}%
\bibitem [{\citenamefont {Motta}\ and\ \citenamefont
  {Sanvito}(2014)}]{doi:10.1021/ct500390a}%
  \BibitemOpen
  \bibfield  {author} {\bibinfo {author} {\bibfnamefont {C.}~\bibnamefont
  {Motta}}\ and\ \bibinfo {author} {\bibfnamefont {S.}~\bibnamefont
  {Sanvito}},\ }\href {\doibase 10.1021/ct500390a} {\bibfield  {journal}
  {\bibinfo  {journal} {J. Chem. Theo. Comp.}\ }\textbf {\bibinfo {volume}
  {10}},\ \bibinfo {pages} {4624} (\bibinfo {year} {2014})}\BibitemShut
  {NoStop}%
\bibitem [{\citenamefont {Wannier}(1937)}]{PhysRev.52.191}%
  \BibitemOpen
  \bibfield  {author} {\bibinfo {author} {\bibfnamefont {G.~H.}\ \bibnamefont
  {Wannier}},\ }\href {\doibase 10.1103/PhysRev.52.191} {\bibfield  {journal}
  {\bibinfo  {journal} {Phys. Rev.}\ }\textbf {\bibinfo {volume} {52}},\
  \bibinfo {pages} {191} (\bibinfo {year} {1937})}\BibitemShut {NoStop}%
\bibitem [{\citenamefont {Wannier}(1962)}]{RevModPhys.34.645}%
  \BibitemOpen
  \bibfield  {author} {\bibinfo {author} {\bibfnamefont {G.~H.}\ \bibnamefont
  {Wannier}},\ }\href {\doibase 10.1103/RevModPhys.34.645} {\bibfield
  {journal} {\bibinfo  {journal} {Rev. Mod. Phys.}\ }\textbf {\bibinfo {volume}
  {34}},\ \bibinfo {pages} {645} (\bibinfo {year} {1962})}\BibitemShut
  {NoStop}%
\bibitem [{\citenamefont {Marzari}\ and\ \citenamefont
  {Vanderbilt}(1997)}]{PhysRevB.56.12847}%
  \BibitemOpen
  \bibfield  {author} {\bibinfo {author} {\bibfnamefont {N.}~\bibnamefont
  {Marzari}}\ and\ \bibinfo {author} {\bibfnamefont {D.}~\bibnamefont
  {Vanderbilt}},\ }\href {\doibase 10.1103/PhysRevB.56.12847} {\bibfield
  {journal} {\bibinfo  {journal} {Phys. Rev. B}\ }\textbf {\bibinfo {volume}
  {56}},\ \bibinfo {pages} {12847} (\bibinfo {year} {1997})}\BibitemShut
  {NoStop}%
\bibitem [{\citenamefont {Marzari}\ \emph {et~al.}(2012)\citenamefont
  {Marzari}, \citenamefont {Mostofi}, \citenamefont {Yates}, \citenamefont
  {Souza},\ and\ \citenamefont {Vanderbilt}}]{RevModPhys.84.1419}%
  \BibitemOpen
  \bibfield  {author} {\bibinfo {author} {\bibfnamefont {N.}~\bibnamefont
  {Marzari}}, \bibinfo {author} {\bibfnamefont {A.~A.}\ \bibnamefont
  {Mostofi}}, \bibinfo {author} {\bibfnamefont {J.~R.}\ \bibnamefont {Yates}},
  \bibinfo {author} {\bibfnamefont {I.}~\bibnamefont {Souza}}, \ and\ \bibinfo
  {author} {\bibfnamefont {D.}~\bibnamefont {Vanderbilt}},\ }\href {\doibase
  10.1103/RevModPhys.84.1419} {\bibfield  {journal} {\bibinfo  {journal} {Rev.
  Mod. Phys.}\ }\textbf {\bibinfo {volume} {84}},\ \bibinfo {pages} {1419}
  (\bibinfo {year} {2012})}\BibitemShut {NoStop}%
\bibitem [{\citenamefont {Soler}\ \emph {et~al.}(2002)\citenamefont {Soler},
  \citenamefont {Artacho}, \citenamefont {Gale}, \citenamefont {Garc{\'\i}a},
  \citenamefont {Junquera}, \citenamefont {Ordej{\'o}n},\ and\ \citenamefont
  {S{\'a}nchez-Portal}}]{0953-8984-14-11-302}%
  \BibitemOpen
  \bibfield  {author} {\bibinfo {author} {\bibfnamefont {J.~M.}\ \bibnamefont
  {Soler}}, \bibinfo {author} {\bibfnamefont {E.}~\bibnamefont {Artacho}},
  \bibinfo {author} {\bibfnamefont {J.~D.}\ \bibnamefont {Gale}}, \bibinfo
  {author} {\bibfnamefont {A.}~\bibnamefont {Garc{\'\i}a}}, \bibinfo {author}
  {\bibfnamefont {J.}~\bibnamefont {Junquera}}, \bibinfo {author}
  {\bibfnamefont {P.}~\bibnamefont {Ordej{\'o}n}}, \ and\ \bibinfo {author}
  {\bibfnamefont {D.}~\bibnamefont {S{\'a}nchez-Portal}},\ }\href
  {http://stacks.iop.org/0953-8984/14/i=11/a=302} {\bibfield  {journal}
  {\bibinfo  {journal} {J. Phys. Condens. Matter}\ }\textbf {\bibinfo {volume}
  {14}},\ \bibinfo {pages} {2745} (\bibinfo {year} {2002})}\BibitemShut
  {NoStop}%
\bibitem [{\citenamefont {Mostofi}\ \emph {et~al.}(2008)\citenamefont
  {Mostofi}, \citenamefont {Yates}, \citenamefont {Lee}, \citenamefont {Souza},
  \citenamefont {Vanderbilt},\ and\ \citenamefont {Marzari}}]{Mostofi2008685}%
  \BibitemOpen
  \bibfield  {author} {\bibinfo {author} {\bibfnamefont {A.~A.}\ \bibnamefont
  {Mostofi}}, \bibinfo {author} {\bibfnamefont {J.~R.}\ \bibnamefont {Yates}},
  \bibinfo {author} {\bibfnamefont {Y.-S.}\ \bibnamefont {Lee}}, \bibinfo
  {author} {\bibfnamefont {I.}~\bibnamefont {Souza}}, \bibinfo {author}
  {\bibfnamefont {D.}~\bibnamefont {Vanderbilt}}, \ and\ \bibinfo {author}
  {\bibfnamefont {N.}~\bibnamefont {Marzari}},\ }\href {\doibase
  http://dx.doi.org/10.1016/j.cpc.2007.11.016} {\bibfield  {journal} {\bibinfo
  {journal} {Comput. Phys. Commun.}\ }\textbf {\bibinfo {volume} {178}},\
  \bibinfo {pages} {685 } (\bibinfo {year} {2008})}\BibitemShut {NoStop}%
\bibitem [{\citenamefont {Ning}\ and\ \citenamefont {Tian}(2009)}]{B908802D}%
  \BibitemOpen
  \bibfield  {author} {\bibinfo {author} {\bibfnamefont {Z.}~\bibnamefont
  {Ning}}\ and\ \bibinfo {author} {\bibfnamefont {H.}~\bibnamefont {Tian}},\
  }\href {\doibase 10.1039/B908802D} {\bibfield  {journal} {\bibinfo  {journal}
  {Chem. Commun.}\ }\textbf {\bibinfo {volume} {37}},\ \bibinfo {pages} {5483}
  (\bibinfo {year} {2009})}\BibitemShut {NoStop}%
\bibitem [{\citenamefont {Fern{\'a}ndez-Seivane}\ \emph
  {et~al.}(2006)\citenamefont {Fern{\'a}ndez-Seivane}, \citenamefont
  {Oliveira}, \citenamefont {Sanvito},\ and\ \citenamefont
  {Ferrer}}]{0953-8984-18-34-012}%
  \BibitemOpen
  \bibfield  {author} {\bibinfo {author} {\bibfnamefont {L.}~\bibnamefont
  {Fern{\'a}ndez-Seivane}}, \bibinfo {author} {\bibfnamefont {M.~A.}\
  \bibnamefont {Oliveira}}, \bibinfo {author} {\bibfnamefont {S.}~\bibnamefont
  {Sanvito}}, \ and\ \bibinfo {author} {\bibfnamefont {J.}~\bibnamefont
  {Ferrer}},\ }\href {http://stacks.iop.org/0953-8984/18/i=34/a=012} {\bibfield
   {journal} {\bibinfo  {journal} {J. Phys. Condens. Matter}\ }\textbf
  {\bibinfo {volume} {18}},\ \bibinfo {pages} {7999} (\bibinfo {year}
  {2006})}\BibitemShut {NoStop}%
\bibitem [{Note1()}]{Note1}%
  \BibitemOpen
  \bibinfo {note} {The correctness of the elements $U_{pm}^{s}(\protect \mathbf
  {k})$ and $e^{i\protect \mathbf {k}\cdot \protect \mathbf {R}}$ is easily
  verified by ensuring that the following relation is satisfied \begin
  {multline}\label {equ_footnote} \begin {split} \mathinner {\delimiter
  "426830A {w_{m\protect \mathbf {R}_1}|w_{n\protect \mathbf {R}_2}}\delimiter
  "526930B }&=\protect \frac {1}{N}\DOTSB \sum@ \slimits@ _p \DOTSI \intop
  \ilimits@ _\protect \mathrm {FBZ} d\protect \mathbf {k} \mathinner
  {\delimiter "426830A {w_{m\protect \mathbf {R}_1}|\psi _{p\protect \mathbf
  {k}}}\delimiter "526930B }\mathinner {\delimiter "426830A {\psi _{p\protect
  \mathbf {k}}|w_{n\protect \mathbf {R}_2}}\delimiter "526930B }=\\ &=\protect
  \frac {1}{N}\DOTSB \sum@ \slimits@ _p \DOTSI \intop \ilimits@ _\protect
  \mathrm {FBZ} d\protect \mathbf {k}U^*_{pm}(\protect \mathbf
  {k})U_{pn}(\protect \mathbf {k})e^{i\protect \mathbf {k}\cdot (\protect
  \mathbf {R}_1-\protect \mathbf {R}_2)}=\\ &=\delta _{m,n}\delta _{\protect
  \mathbf {R}_1,\protect \mathbf {R}_2} \protect \tmspace +\medmuskip
  {.2222em}. \end {split} \end {multline}}\BibitemShut {NoStop}%
\bibitem [{Note2()}]{Note2}%
  \BibitemOpen
  \bibinfo {note} {The real spherical harmonics are constructed from the
  complex ones, $\mathinner {|{l,m}\delimiter "526930B }$, as $\mathinner
  {|{l,M}\delimiter "526930B }=\protect \frac {1}{\protect \sqrt 2}[\mathinner
  {|{l,m}\delimiter "526930B }+(-1)^m\mathinner {|{l,-m}\delimiter "526930B }]$
  and $\mathinner {|{l,-M}\delimiter "526930B }=\protect \frac {1}{i\protect
  \sqrt 2}[\mathinner {|{l,m}\delimiter "526930B }-(-1)^m\mathinner
  {|{l,-m}\delimiter "526930B }]$. For $M=0$ the real and complex spherical
  harmonics coincide.}\BibitemShut {Stop}%
\bibitem [{\citenamefont {Romanov}(1993)}]{PbZeo}%
  \BibitemOpen
  \bibfield  {author} {\bibinfo {author} {\bibfnamefont {S.}~\bibnamefont
  {Romanov}},\ }\href {http://stacks.iop.org/0953-8984/5/i=8/a=009} {\bibfield
  {journal} {\bibinfo  {journal} {Journal of Physics: Condensed Matter}\
  }\textbf {\bibinfo {volume} {5}},\ \bibinfo {pages} {1081} (\bibinfo {year}
  {1993})}\BibitemShut {NoStop}%
\bibitem [{\citenamefont {Moulin}\ \emph {et~al.}(2010)\citenamefont {Moulin},
  \citenamefont {Niess}, \citenamefont {Maaloum}, \citenamefont {Buhler},
  \citenamefont {Nyrkova},\ and\ \citenamefont
  {Giuseppone}}]{ANIE:ANIE201001833}%
  \BibitemOpen
  \bibfield  {author} {\bibinfo {author} {\bibfnamefont {E.}~\bibnamefont
  {Moulin}}, \bibinfo {author} {\bibfnamefont {F.}~\bibnamefont {Niess}},
  \bibinfo {author} {\bibfnamefont {M.}~\bibnamefont {Maaloum}}, \bibinfo
  {author} {\bibfnamefont {E.}~\bibnamefont {Buhler}}, \bibinfo {author}
  {\bibfnamefont {I.}~\bibnamefont {Nyrkova}}, \ and\ \bibinfo {author}
  {\bibfnamefont {N.}~\bibnamefont {Giuseppone}},\ }\href {\doibase
  10.1002/anie.201001833} {\bibfield  {journal} {\bibinfo  {journal} {Angew.
  Chem. Int. Ed.}\ }\textbf {\bibinfo {volume} {49}},\ \bibinfo {pages} {6974}
  (\bibinfo {year} {2010})}\BibitemShut {NoStop}%
\bibitem [{\citenamefont {Faramarzi}\ \emph {et~al.}(2012)\citenamefont
  {Faramarzi}, \citenamefont {Niess}, \citenamefont {Moulin}, \citenamefont
  {Maaloum}, \citenamefont {Dayen}, \citenamefont {Beaufrand}, \citenamefont
  {Zanettini}, \citenamefont {Doudin},\ and\ \citenamefont
  {Giuseppone}}]{Vina}%
  \BibitemOpen
  \bibfield  {author} {\bibinfo {author} {\bibfnamefont {V.}~\bibnamefont
  {Faramarzi}}, \bibinfo {author} {\bibfnamefont {F.}~\bibnamefont {Niess}},
  \bibinfo {author} {\bibfnamefont {E.}~\bibnamefont {Moulin}}, \bibinfo
  {author} {\bibfnamefont {M.}~\bibnamefont {Maaloum}}, \bibinfo {author}
  {\bibfnamefont {J.-F.}\ \bibnamefont {Dayen}}, \bibinfo {author}
  {\bibfnamefont {J.-B.}\ \bibnamefont {Beaufrand}}, \bibinfo {author}
  {\bibfnamefont {S.}~\bibnamefont {Zanettini}}, \bibinfo {author}
  {\bibfnamefont {B.}~\bibnamefont {Doudin}}, \ and\ \bibinfo {author}
  {\bibfnamefont {N.}~\bibnamefont {Giuseppone}},\ }\href@noop {} {\bibfield
  {journal} {\bibinfo  {journal} {Nature Chemistry}\ }\textbf {\bibinfo
  {volume} {4}},\ \bibinfo {pages} {485} (\bibinfo {year} {2012})}\BibitemShut
  {NoStop}%
\bibitem [{\citenamefont {Akande}\ \emph {et~al.}(2014)\citenamefont {Akande},
  \citenamefont {Bhattacharya}, \citenamefont {Cathcart},\ and\ \citenamefont
  {Sanvito}}]{Akin}%
  \BibitemOpen
  \bibfield  {author} {\bibinfo {author} {\bibfnamefont {A.}~\bibnamefont
  {Akande}}, \bibinfo {author} {\bibfnamefont {S.}~\bibnamefont
  {Bhattacharya}}, \bibinfo {author} {\bibfnamefont {T.}~\bibnamefont
  {Cathcart}}, \ and\ \bibinfo {author} {\bibfnamefont {S.}~\bibnamefont
  {Sanvito}},\ }\href@noop {} {\bibfield  {journal} {\bibinfo  {journal} {J.
  Chem. Phys}\ }\textbf {\bibinfo {volume} {140}},\ \bibinfo {eid} {074301}
  (\bibinfo {year} {2014})}\BibitemShut {NoStop}%
\bibitem [{\citenamefont {Bhattacharya}\ \emph {et~al.}(2014)\citenamefont
  {Bhattacharya}, \citenamefont {Akande},\ and\ \citenamefont
  {Sanvito}}]{C4CC01710B}%
  \BibitemOpen
  \bibfield  {author} {\bibinfo {author} {\bibfnamefont {S.}~\bibnamefont
  {Bhattacharya}}, \bibinfo {author} {\bibfnamefont {A.}~\bibnamefont
  {Akande}}, \ and\ \bibinfo {author} {\bibfnamefont {S.}~\bibnamefont
  {Sanvito}},\ }\href {\doibase 10.1039/C4CC01710B} {\bibfield  {journal}
  {\bibinfo  {journal} {Chem. Commun.}\ }\textbf {\bibinfo {volume} {50}},\
  \bibinfo {pages} {6626} (\bibinfo {year} {2014})}\BibitemShut {NoStop}%
\end{thebibliography}
\end{document}